# Formulation and numerical implementation of micro-scale boundary conditions for particle aggregates

J. Liu, E. Bosco, A.S.J. Suiker[*]

Department of the Built Environment,
Eindhoven University of Technology, The Netherlands

February 13, 2017

**Abstract**

Novel numerical algorithms are presented for the implementation of micro-scale boundary conditions of particle aggregates modelled with the discrete element method. The algorithms are based on a servo-control methodology, using a feedback principle comparable to that of algorithms commonly applied within control theory of dynamic systems. The boundary conditions are defined in accordance with the large deformation theory, and are imposed on a frame of boundary particles surrounding the interior granular micro-structure. Following the formulation presented in Miehe et al., (2010), *Int. J. Num. Meth. Engng.* **83**, pp. 1206-1236, first three types of classical boundary conditions are considered, in accordance with i) a homogeneous deformation and zero particle rotation (D), ii) a periodic particle displacement and rotation (P), and iii) a uniform particle force and free particle rotation (T). The algorithms can be straightforwardly combined with commercially available discrete element codes, thereby enabling the determination of the solution of boundary-value problems at the micro-scale only, or at multiple scales via a micro-to-macro coupling with a finite element model. The performance of the algorithms is tested by means of discrete element method simulations on regular monodisperse packings and irregular polydisperse packings composed of frictional particles, which were subjected to various loading paths. The simulations provide responses with the typical stiff and soft bounds for the (D) and (T) boundary conditions, respectively, and illustrate for the (P) boundary condition a relatively fast convergence of the apparent macroscopic properties under an increasing packing size. Finally, a homogenization framework is derived for the implementation of mixed (D)-(P)-(T) boundary conditions that satisfy the Hill-Mandel micro-heterogeneity condition on energy consistency at the micro- and macro-scales of the granular system. The numerical algorithm for the mixed boundary conditions is developed and tested for the case of an infinite layer subjected to a vertical compressive stress and a horizontal shear deformation, whereby the response computed for a layer of cohesive particles is compared against that for a layer of frictional particles.

## 1 Introduction

The accurate computation of the non-linear failure and deformation behavior of heterogeneous granular systems commonly requires a resolution of the complex mechanical interactions and deformation mechanisms at the particle scale, which can be adequately accounted for by using the discrete element method (DEM), see e.g., [1, 2, 3, 4, 5, 6, 7, 8, 9, 10] and references therein. For practical granular systems composed of a vast number of particles, however, it is infeasible to simulate each particle as an individual discrete object, since this leads to DEM models with an enormously large number of degrees of freedom, and consequently, to impractical computation times. To circumvent this problem, advanced multi-scale frameworks have been developed, where the mechanical responses at the particle micro-scale and the structural macro-scale are hierarchically coupled in an computationally economical fashion. This is accomplished by simulating the macro-scale problem under consideration with the finite element method (FEM), whereby in every integration point the response to the corresponding deformation is calculated by means of a DEM model that accurately and efficiently represents the complex particle behavior at the micro-scale. Examples of coupled FEM-DEM approaches for granular materials can be found in [11, 12, 13, 14, 15], illustrating the use of various averaging theorems for relating force and displacement measures at the particle micro-scale to stress and strain measures at the structural macro-scale. Specific aspects that should deserve more attention in FEM-DEM homogenization methods, but often are neglected for reasons of simplicity, refer to i) the Hill-Mandel micro-heterogeneity condition, which enforces consistency of energy at the micro- and macro-scales, ii) the effect of particle rotations in the formulation of micro-to-macro scale-transitions, and iii) a rigorous generalization of the multi-scale approach within the theory of large deformations.

---

[*]Correspondence to: a.s.j.suiker@tue.nl, Department of the Built Environment, Eindhoven University of Technology, P.O. Box 513, 5600 MB Eindhoven, The Netherlands.

The computational homogenization framework presented in [16] does include the three aspects mentioned above, and calculates the micro-scale response of a granular packing with a DEM model equipped with a frame of boundary particles at which the finite deformation following from the macro-scale is imposed. The formulation considers three types of micro-scale boundary conditions for the boundary particles, namely i) homogeneous deformation and zero particle rotation (D), ii) periodic particle displacements and rotations (P), and iii) uniform particle force and free particle rotation (T), where the abbreviations (D), (P) and (T) are adopted from analogous, classical boundary conditions used in continuum homogenization theories, referring to the displacement, periodic and traction boundary conditions, respectively. The numerical implementation of the boundary conditions in [16] is done via a penalty method, where the violation of the boundary conditions is punished by increasing the total virtual work of the particle packing, through the introduction of additional forces and moments on the frame of boundary particles. Although the algorithm presented in [16] has been nicely generalized for the three types of boundary conditions in a mathematically elegant and transparent fashion, due to the nature of the penalty method the expression for the homogenized stress of the particle packing becomes explicitly dependent on the value of the penalty parameter, and thereby looses its physical interpretation. In addition, in DEM models the penalty parameter may be difficult to control and must be chosen sufficiently large in order for the penalty function to be effective, which may induce numerical instabilities [17, 18]. Another characteristic of the penalty method is that it requires the constraint equations to be satisfied "approximately" instead of "exactly", whereby the accuracy of the approximation is determined by the magnitude of the penalty parameter. As a consequence, the limit case at which the boundary conditions are met exactly is not rigorously retrieved from the formulation, since the homogenized stress of the particle packing then vanishes, see expression (44) in [16].

In order to improve on the algorithmic drawbacks mentioned above, in the present communication an alternative numerical algorithm is proposed for the implementation of the homogenization framework presented in [16]. This algorithm is based on a servo-control methodology, using a feedback principle comparable to that of algorithms commonly applied within control theory of dynamic systems [19]. Accordingly, the displacements and rotations of the particles of the boundary frame are iteratively adapted from a gradually diminishing discrepancy between the measured and desired values of the micro-scale boundary condition. A strong point of this approach is that it is relatively simple to implement, and only affects the interface communicating information between the macro-scale FEM and micro-scale DEM models. In other words, it does not require internal modifications of the FEM and DEM source codes, so that the approach can also be combined with commercially available software for which the user typically has no access to the source code. In addition to its simplicity, the servo-control algorithm preserves the physical interpretation of the homogenized stress measure derived for the particle packing, and furnishes a realistic value for the stress in the limit case at which the micro-scale boundary condition is met exactly. It is noted that the algorithm only considers the (P) and (T) boundary conditions, since for a macro-scale problem discretized with a displacement-based FEM code, the (D) boundary condition can be implemented in a straightforward fashion, without the use of iterations.

Apart from providing servo-control algorithms for the individual (P) and (T) boundary conditions, a novel formulation for *mixed (D)-(P)-(T) boundary conditions* is derived, and subsequently cast into a numerical formalism. The formulation is proven to satisfy the Hill-Mandel micro-heterogeneity condition, and therefore is very useful for i) a consistent derivation of macro-scale constitutive relations from standard material tests on particle aggregates subjected to any combination of (D)-, (P)- and/or (T)-type boundary conditions, and ii) the efficient computation of the homogenized response of large-scale particle aggregates characterized by a spatial periodicity in one or two directions, i.e., granular layers exposed to uniform (D) and/or (T) boundary conditions at their top and bottom surfaces. It will be demonstrated that the formulation allows to impose the (D) and (T) boundary conditions both at separate and identical parts of the layer boundary, where in the latter case the (D) and (T) contributions obviously need to be applied along different orthonormal directions.

The performance of the servo-control algorithms developed for the various micro-scale boundary conditions is tested by using monodisperse and polydisperse frictional and cohesive packings composed of two-dimensional, circular particles and subjected to various loading paths. These examples illustrate the basic features of each of the boundary conditions in full detail. Despite the focus on two-dimensional particle systems, it should be mentioned that the extension of the present framework towards three-dimensional granular systems is trivial, and can be made without the introduction of additional prerequisites.

The paper is organized as follows. Section 2 presents a review of the numerical homogenization framework for particle aggregates, and outlines the formulations of the micro-scale (D), (P) and (T) boundary conditions proposed in [16]. Section 3 discusses the numerical implementation of the micro-scale boundary conditions, where for the (P) and (T) boundary conditions two different servo-control algorithms are presented, which include or not an initial prediction of the displacements of the boundary particles based on their positions calculated at the previous loading step. In Section 4 the performance of the numerical algorithms is tested on monodisperse and polydisperse frictional packings subjected to various loading paths. The numerical results clearly illustrate the characteristic differences in response for the three types of boundary conditions, and show their response convergence behavior under increasing sample size. Section 5 presents the formulation for the mixed boundary conditions, and provides the details of the servo-control algorithm and its numerical performance for the cases of infinite frictional and cohesive granular layers loaded by a vertical compressive stress, and subsequently subjected to a horizontal shear deformation. Some concluding remarks are provided in Section 6.

In terms of notation, the cross product and dyadic product of two vectors are, respectively, designated as $\mathbf{a} \times \mathbf{b} = e_{ijk}a_i b_j \mathbf{e}_k$ and $\mathbf{a} \otimes \mathbf{b} = a_i b_j \mathbf{e}_i \otimes \mathbf{e}_j$, where $e_{ijk}$ is the permutation symbol, $\mathbf{e}_i$, $\mathbf{e}_j$ and $\mathbf{e}_k$ are unit vectors in a Cartesian vector basis, and Einstein's summation convention is used on repeated tensor indices. The inner product between two vectors is given by $\mathbf{a} \cdot \mathbf{b} = a_i b_i$, and between two second-order tensors by $\mathbf{A} : \mathbf{B} = A_{ij} B_{ij}$. The action of a second-order tensor on a vector is indicated as $\mathbf{A} \cdot \mathbf{b} = A_{ij} b_j \mathbf{e}_i$. The superscript $T$ is used to indicate the transpose of a vector or a tensor. Further, $\mathbf{I} = \delta_{ij} \mathbf{e}_i \otimes \mathbf{e}_j$ denotes the second-order identity tensor, with $\delta_{ij}$ the Kronecker delta symbol.

Since the present study focuses on two-dimensional particle aggregates, throughout the paper the dimensions related to volume, area, stress and mass density are consistently presented in their reduced form as length$^2$, length, force/length and mass/length$^2$, respectively.

## 2 Micro-macro transitions for particle aggregates

### 2.1 Micro-scale geometry

The initial micro-scale granular system is characterized by a two-dimensional square domain of $P + Q$ rigid particles, which are partitioned into $P$ inner particles $\mathcal{P}_p$, with $p = 1, .., P$, and $Q$ boundary particles $\mathcal{P}_q$, with $q = 1, .., Q$, colored in yellow and red in Figure 1(a), respectively. The boundary particles can be further split into corner particles $\mathcal{P}_c$ with $c = 1, .., 4$ and the remaining edge particles $\mathcal{P}_e$ with $e = 1, .., E = Q - 4$. The initial interior domain $V$ comprises the inner particles with their center points as $\mathbf{X}_p \in \mathcal{P}_p$ with $p = 1, .., P$. The boundary $\partial V$ is defined by the boundary particles, whose center points in the initial configuration are $\mathbf{X}_q \in \mathcal{P}_q$ with $q = 1, .., Q$. The macroscopic deformation of the granular microstructure is imposed via the frame of boundary particles $\mathcal{P}_q$, as a result of which the center points of the inner and boundary particles become located at $\mathbf{x}_p$ and $\mathbf{x}_q$, respectively, see Figure 1(b). In the current configuration, the boundary particles $\mathcal{P}_q$ are subjected to boundary forces $\mathbf{a}_q$, boundary moments $\mathbf{m}_q$, and particle contact forces $\mathbf{f}_q^c$, see Figure 1(c), while the inner particles $\mathcal{P}_p$ are subjected to particle contact forces $\mathbf{f}_p^c$, see Figure 1(d), with the superscript $c$ denoting a contact with a neighbour particle.

The macroscopic response of a granular assembly is derived by transforming relevant principles used in first-order homogenization theories [20, 21, 22, 23] from a continuous setting to a discrete setting. Accordingly, at the centroids of the boundary particles $\mathcal{P}_q$ the finite area vectors $\mathbf{A}_q$ and forces $\mathbf{a}_q$ are derived from infinitesimal area vectors and forces, respectively,

$$\int_{\partial V} \mathbf{N} ds \to \mathbf{A}_q \quad \text{and} \quad \int_{\partial V} \mathbf{t}\, ds \to \mathbf{a}_q \quad \text{for} \quad q = 1 ..., Q\,, \tag{1}$$

with $\mathbf{N}$ the vector pointing in the outward normal direction of the boundary $\partial V$ of the initial particle volume $V$, $\mathbf{t}$ being the boundary traction, and $ds$ indicating an infinitesimal part of the boundary surface. Various expressions for $\mathbf{A}_q$ have been presented in the literature, see e.g., [24, 25]. In the present communication the initial area vector is computed by accounting for the different radii of the boundary particles:

$$\mathbf{A}_q = \frac{R_q}{R_q + R_{q-1}}(\mathbf{X}_q - \mathbf{X}_{q-1}) \times \mathbf{e}_3 + \frac{R_q}{R_q + R_{q+1}}(\mathbf{X}_{q+1} - \mathbf{X}_q) \times \mathbf{e}_3\,, \tag{2}$$

where $R_{q+1}$, $R_q$ and $R_{q-1}$ are the radii of adjacent boundary particles $q + 1$, $q$ and $q - 1$, respectively. Further, $\mathbf{e}_3$ represents the unit vector in the out-of-plane direction of the two-dimensional particle structure, see also Figure 1(a). Note that in (2) the boundary particles must be numbered in the anti-clockwise direction in order to obtain an area vector pointing in the outward normal direction of the boundary.

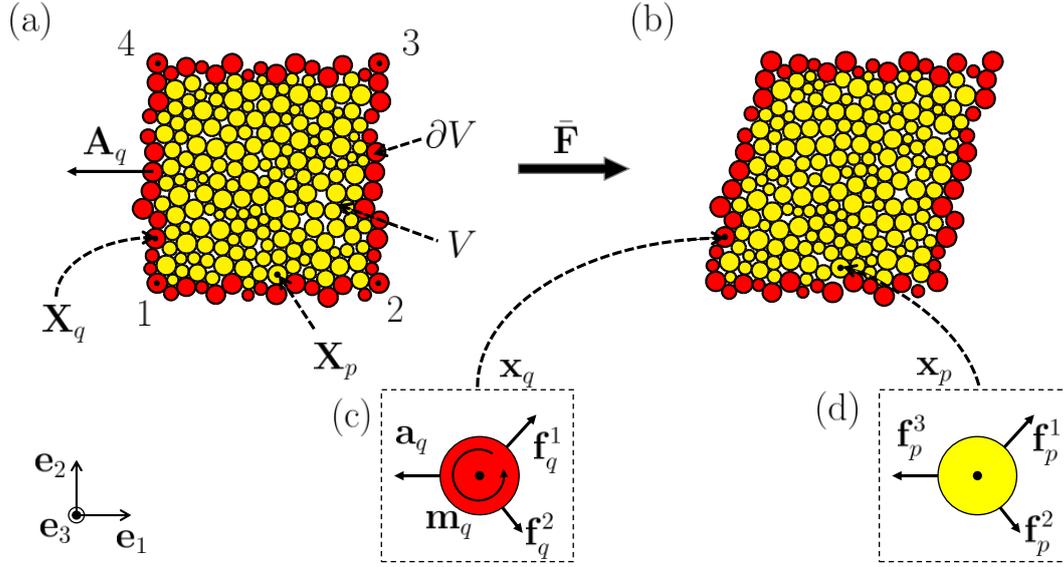

Figure 1: (a) Two-dimensional particle aggregate of initial volume $V$ and boundary $\partial V$. Yellow and red colors refer to inner $\mathcal{P}_p$ and boundary $\mathcal{P}_q$ particles, respectively; (b) Particle aggregate in the current configuration; (c) Boundary forces $\mathbf{a}_q$, boundary moments $\mathbf{m}_q$, and particle contact forces $\mathbf{f}_q^c$ acting on the boundary particles $\mathcal{P}_q$; (d) Particle contact forces $\mathbf{f}_p^c$ acting on the inner particles $\mathcal{P}_p$.

## 2.2 Micro-scale governing equations

### 2.2.1 Equilibrium conditions

In the absence of body forces, the mechanical equilibrium of a granular microstructure can be formulated in terms of the boundary forces $\mathbf{a}_q$ and moments $\mathbf{m}_q$ acting on the boundary particles $\mathcal{P}_q$:

$$\sum_{q=1}^{Q} \mathbf{a}_q = \mathbf{0} \quad \text{and} \quad \sum_{q=1}^{Q} (\mathbf{x}_q \times \mathbf{a}_q + \mathbf{m}_q) = \mathbf{0} \quad \text{for} \quad q = 1..., Q, \tag{3}$$

with $\mathbf{x}_q$ the current position vector of the boundary particles. The boundary forces $\mathbf{a}_q$ and moments $\mathbf{m}_q$ thus drive the overall, macroscopic deformation of the granular system via the frame of boundary particles.

Note that, besides global equilibrium (3), local equilibrium conditions may be formulated for each of the inner particles $\mathcal{P}_p$, which interact through contact forces $\mathbf{f}_p^c$ at discrete contact points $\mathbf{x}_p^c$ on the particle surfaces:

$$\sum_{c=1}^{N_p^c} \mathbf{f}_p^c = \mathbf{0} \quad \text{and} \quad \sum_{c=1}^{N_p^c} (\mathbf{x}_p^c - \mathbf{x}_p) \times \mathbf{f}_p^c = \mathbf{0} \quad \text{for} \quad p = 1..., P, \tag{4}$$

with the superscript $c$ referring to a particle contact, $N_p^c$ being the number of contact forces related to particle $p$ and $\mathbf{x}_p$ is the current position vector of the inner particle. Analogous conditions may be written for the boundary particles $\mathcal{P}_q$, for which discrete contact forces $\mathbf{f}_q^c$ act at contact points $\mathbf{x}_q^c$ on the particle surfaces. The frame of boundary particles is driven by boundary forces $\mathbf{a}_q$ and boundary moments $\mathbf{m}_q$,

$$\sum_{c=1}^{N_q^c} \mathbf{f}_q^c = -\mathbf{a}_q \quad \text{and} \quad \sum_{q=1}^{N_q^c} (\mathbf{x}_q^c - \mathbf{x}_q) \times \mathbf{f}_q^c = -\mathbf{m}_q \quad \text{for} \quad q = 1..., Q, \tag{5}$$

where $N_q^c$ is the number of contact forces for particle $q$. Note that the combination of expressions (4) and (5) is in correspondence with relation (3).

### 2.2.2 Particle contact laws

In order to solve the micro-scale problem, the constitutive response of the granular assembly needs to be defined through a relation between the contact forces $\mathbf{f}_i^c$ (or contact moments $\mathbf{m}_i^c$), with $i = 1, ..., P + Q$, and the corresponding contact displacements $\Delta \mathbf{u}_i^c$ (or contact rotations $\Delta \boldsymbol{\theta}_i^c$). For the sake of clarity, in the following the superscript $c$ and subscript $i$ will be dropped. Two types of particle contact interactions will be considered, which are referred to as *frictional contact* and *cohesive contact*.

In accordance with [1], in the *frictional contact* law the normal particle contact force $f_n$ is proportional to the normal overlap $\Delta u_n$ between two particles in contact via a multiplication by the normal contact stiffness $k_n$. The tangential particle contact force $f_s$ is proportional to the relative tangential displacement $\Delta u_s$ at the particle contact via a multiplication by the tangential contact stiffness $k_s$, up to a limit value at which frictional sliding starts, as defined by the normal force multiplied by the friction coefficient $\mu$. The contact law is thus expressed as

$$f_n = k_n \Delta u_n \quad \text{and} \quad f_s = \begin{cases} k_s \Delta u_s & \text{if} \quad f_s < \mu f_n, \\ \mu f_n & \text{otherwise} \end{cases}. \quad (6)$$

For the *cohesive contact* law, two particles in contact are assumed to be initially bonded according to the constitutive model presented in [26, 27], which proposes a linear relation between the force (or moment) and the corresponding relative displacement (or rotation) at the particle contact:

$$f_n^b = k_n^b \Delta u_n \quad \text{and} \quad f_s^b = k_s^b \Delta u_s \quad \text{and} \quad m_\theta^b = k_\theta^b \Delta \theta^b, \quad (7)$$

where the superscript $b$ refers to "bond". In specific, the relative displacements between two particles in the normal and tangential directions of the contact, $\Delta u_n$ and $\Delta u_s$, are, respectively, related to the normal and tangential bond forces $f_n^b$ and $f_s^b$ through a multiplication by the bond stiffnesses $k_n^b$ and $k_s^b$, respectively. Similarly, the relative angular rotation $\Delta \theta^b$ is related to the contact moment $m_\theta^b$ through a multiplication by the bond bending stiffness $k_\theta^b$. The bond between two particles is considered as broken when the following failure criterion is met:

$$\frac{f_n^b}{f_n^{b,u}} + \frac{f_s^b}{f_s^{b,u}} + \frac{m_\theta^b}{m_\theta^{b,u}} = 1, \quad (8)$$

where $f_n^{b,u}$ is the (ultimate) tensile strength, $f_s^{b,u}$ is the shear strength and $m_\theta^{b,u}$ is the bending strength. After breakage of the contact the particle interaction is described by the frictional contact law presented in expression (6).

### 2.2.3 Dynamic relaxation

The equilibrium conditions described by equations (3) to (5) are solved by applying a dynamic relaxation method, in which the kinetic energy activated by the applied deformation is dissipated to arrive at the equilibrium state. For each particle $i$, where $i = 1, .., P + Q$, a vector of generalized coordinates is defined as $\mathbf{d}_i = [\mathbf{x}_i, \boldsymbol{\theta}_i \cdot \mathbf{e}_3]^T$, which includes the particle center location $\mathbf{x}_i$ and rotation $\boldsymbol{\theta}_i$. In addition, a generalized force vector is introduced, $\mathbf{p}_i = [\mathbf{f}_i, \mathbf{m}_i \cdot \mathbf{e}_3]^T$, which contains the forces and moments acting on the particle. Accordingly, the generalized equation of motion of particle $i$ can be expressed as

$$\mathbf{M}_i \ddot{\mathbf{d}}_i = (\mathbf{p}_r + \mathbf{p}_d)_i \quad \text{for} \quad i = 1..., P + Q, \quad (9)$$

where the mass matrix $\mathbf{M}_i = \text{diag}[M_i, I_i]$ includes the particle mass $M_i$ and particle mass moment of inertia $I_i = 1/2\, M_i R_i^2$, with $R_i$ the particle radius. The term $\ddot{\mathbf{d}}_i$ represents the generalized acceleration vector, with a superimposed dot indicating a derivative with respect to time. The vector $\mathbf{p}_r$ is the generalized force vector composed of the resultant force $\mathbf{f}_r$ and moment $\mathbf{m}_r$ acting on particle $i$, and $\mathbf{p}_d = [\mathbf{f}_d, \mathbf{m}_d \cdot \mathbf{e}_3]^T$ is the vector containing the resulting particle force and moment following from the artificial dissipation applied in the simulations to improve the convergence rate towards the equilibrium state. Following [28], the artificial dissipative force $\boldsymbol{f}_d$ and moment $\boldsymbol{m}_d$ are here defined as

$$\mathbf{f}_d = -\alpha |\mathbf{f}_r| \,\text{sign}(\dot{\mathbf{x}}_i) \quad \text{and} \quad \mathbf{m}_d = -\beta |\mathbf{m}_r| \,\text{sign}(\dot{\boldsymbol{\theta}}_i), \quad (10)$$

where $\alpha$ and $\beta$ are damping values related to (signum functions of) the particle translational velocity $\dot{\mathbf{x}}_i$ and rotational velocity $\dot{\boldsymbol{\theta}}_i$, respectively. Further, $|.|$ refers to the absolute values of the components of the corresponding vector.

The time integration of the governing equations is performed by applying an explicit, first-order finite difference scheme, which, for each time step $t_{h+1}$, with the time increment given by $\Delta t = t_{h+1} - t_h$, allows for an explicit update of the particle acceleration, velocity and displacement, see [29] for more details. The dynamic relaxation process is considered to be converged towards the equilibrium state when the ratio between the kinetic energy $E_k$ of the inner particles in the aggregate and their potential energy $E_p$ is lower than a prescribed tolerance [30], i.e.,

$$E_k/E_p \leq \text{tol}_E, \quad (11)$$

in accordance with the following definitions

$$E_k = \sum_{i=1}^{P} \frac{1}{2} \dot{\mathbf{d}}_i^T \mathbf{M}_i \dot{\mathbf{d}}_i \qquad \text{and} \qquad E_p = \sum_{c=1}^{N^c} \frac{1}{2} \left( k_n (\Delta u_n^c)^2 + k_s (\Delta u_s^c)^2 \right), \qquad (12)$$

where $\Delta u_n^c$ and $\Delta u_s^c$ are the relative displacements in the normal and tangential direction of particle contact $c$ and $N^c$ is the total number of particle contacts. Note that for the cohesive contact law given by equations (7) and (8) the potential energy in (12) needs to be extended with the rotational term $k_\theta (\Delta \theta^c)^2/2$.

Obviously, for deriving the solution of a boundary value problem, the equation of motion (9) and the constitutive response of the particles (6) and (7) should be complemented by the appropriate boundary conditions. As mentioned in the introduction, the numerical implementation of the micro-scale boundary conditions is based on the formulation presented in [16], and the main equations are summarized in Section 2.3 for the sake of clarity.

## 2.3 Micro-scale kinematics and boundary conditions

Consider a rigid particle $i$ within a granular assembly. The current location $\mathbf{x}$ of an arbitrary material point, located within the initial particle volume at $\mathbf{X}$, is defined through the non-linear deformation map $\mathbf{x} = \boldsymbol{\psi}_i(\mathbf{X})$, with $\boldsymbol{\psi}_i$ as

$$\boldsymbol{\psi}_i(\mathbf{X}) = \mathbf{x}_i + \mathbf{Q}_i \cdot (\mathbf{X} - \mathbf{X}_i) \qquad \text{for} \quad i = 1..., P+Q, \qquad (13)$$

where $\mathbf{x}_i$ and $\mathbf{X}_i$ are the current and original positions of the center of particle $i$, and $\mathbf{Q}_i$ is the second-order particle transformation tensor. For plane problems defined with respect to the orthonormal tensor basis $\{\mathbf{e}_k \otimes \mathbf{e}_l\}_{k,l=1,2}^2$, the transformation tensor of particle $i$ can be expressed as $\mathbf{Q}_i = \cos\theta_i \mathbf{e}_1 \otimes \mathbf{e}_1 - \sin\theta_i \mathbf{e}_1 \otimes \mathbf{e}_2 + \sin\theta_i \mathbf{e}_2 \otimes \mathbf{e}_1 + \cos\theta_i \mathbf{e}_2 \otimes \mathbf{e}_2$, with $\theta_i$ the magnitude of the particle center rotation $\boldsymbol{\theta}_i = \theta_i \boldsymbol{e}_3$, where $\boldsymbol{e}_3$ is the unit vector normal to the plane. In addition, the current position of the particle center $\mathbf{x}_i$ can be expressed as the sum of a contribution affine to the macroscopic deformation gradient $\bar{\boldsymbol{F}}$ and a local, micro-scale fluctuation $\mathbf{w}_i$:

$$\mathbf{x}_i = \bar{\boldsymbol{F}} \cdot \mathbf{X}_i + \mathbf{w}_i \qquad \text{for} \quad i = 1..., P+Q. \qquad (14)$$

In homogenization schemes for continuous media, the macro-to-micro scale transition is enforced by requiring the macro-scale deformation gradient to be equal to the volume average of the micro-scale deformation gradient. In a discrete setting, this is equivalent to the condition

$$\bar{\boldsymbol{F}} = \frac{1}{V} \sum_{q=1}^{Q} \mathbf{x}_q \otimes \boldsymbol{A}_q. \qquad (15)$$

Relation (15) can be derived by transforming the volume average of the macro-scale deformation into a surface integral

$$\bar{\boldsymbol{F}} = \frac{1}{V} \int_V \boldsymbol{F} dv = \frac{1}{V} \int_V \nabla \boldsymbol{x} dv = \frac{1}{V} \int_{\partial V} \boldsymbol{x} \otimes \boldsymbol{N} ds, \qquad (16)$$

with $\mathbf{N}$ the vector normal to the outer boundary of the original particle volume, and subsequently performing the transition from a continuous to a discrete setting with the aid of $(1)_1$.

Equation (15) needs to be satisfied by applying specific boundary conditions to the boundary particles of the granular microstructure. For continuous media, this goal is typically accomplished by applying one of the three classical types of boundary conditions, namely i) a homogeneous deformation, also known as the displacement boundary condition and thus abbreviated as (D), ii) periodic displacements (P), and iii) a uniform traction (T), see, e.g., [21, 22, 23]. For discrete particle structures, however, additional conditions need to be imposed on the rotations or moments of the boundary particles. Correspondingly, along the lines of [16], the three boundary conditions mentioned above are extended as i) homogeneous deformation and zero rotation (D), ii) periodic displacement and periodic rotation (P), and iii) uniform force and free rotation (T), of which the formulations are presented below. The abbreviations (D), (P) and (T), although typically used in continuum homogenization theories, are maintained here for reasons of consistency. In addition to the three classical boundary conditions, a novel combination of these boundary conditions has been derived, which will be referred to as "mixed boundary conditions". The corresponding formulation is proven to satisfy the consistency of energy between the microscopic and macroscopic scales of observation, known as the Hill-Mandel micro-heterogeneity condition, and the details are provided in Section 5.

### 2.3.1 Homogeneous deformation and zero rotation (D)

In accordance with this boundary condition, all the boundary particles $\mathcal{P}_q$ are prescribed to have zero micro-scale displacement fluctuations and zero rotations:

$$\mathbf{x}_q = \bar{\boldsymbol{F}} \cdot \boldsymbol{X}_q \quad \text{and} \quad \mathbf{Q}_q = \boldsymbol{I} \quad \text{on} \quad \partial V, \tag{17}$$

where the first expression follows from (14) with the displacement fluctuations as $\mathbf{w}_q = \mathbf{0}$. Due to the second condition in (17) the boundary moments do not vanish, i.e.,

$$\mathbf{m}_q \neq \mathbf{0} \quad \text{on} \quad \partial V. \tag{18}$$

The homogeneous deformation and zero rotation boundary condition is expected to result in a relatively stiff macroscopic response of the particle aggregate.

### 2.3.2 Periodic displacement and periodic rotation (P)

For this boundary condition, both the displacements and rotations of the boundary particles $\mathcal{P}_q$ are related by periodicity requirements:

$$\mathbf{x}_q^+ - \mathbf{x}_q^- = \bar{\boldsymbol{F}} \cdot (\mathbf{X}_q^+ - \mathbf{X}_q^-) \quad \text{and} \quad \mathbf{Q}_q^+ - \mathbf{Q}_q^- = \mathbf{0} \quad \text{on} \quad \partial V, \tag{19}$$

where the superscripts $^+$ and $^-$ refer to corresponding particles on opposite boundaries of the granular assembly. From the viewpoint of equilibrium, the forces and moments on opposite boundaries need to be anti-periodic, thus satisfying the relations

$$\mathbf{a}_q^+ + \mathbf{a}_q^- = \mathbf{0} \quad \text{and} \quad \mathbf{m}_q^+ + \mathbf{m}_q^- = \mathbf{0} \quad \text{on} \quad \partial V. \tag{20}$$

### 2.3.3 Uniform force and free rotation (T)

The boundary forces $\mathbf{a}_q$ of the boundary particles $\mathcal{P}_q$ are here determined from the product of the macroscopic first Piola-Kirchhoff stress $\bar{\boldsymbol{P}}$ and the discrete area vectors $\mathbf{A}_q$ introduced in equation $(1)_1$:

$$\mathbf{a}_q = \bar{\boldsymbol{P}} \cdot \boldsymbol{A}_q \quad \text{on} \quad \partial V. \tag{21}$$

In addition, no constraint is applied to the boundary rotations, so that the boundary moments vanish:

$$\mathbf{m}_q = \mathbf{0} \quad \text{on} \quad \partial V. \tag{22}$$

The uniform force and free rotation boundary condition is expected to provide a relatively soft macroscopic response of the particle aggregate.

## 2.4 Macro-scale stress and Hill-Mandel condition

The first Piola-Kirchhoff stress at the macro scale is defined in terms of the boundary forces $\mathbf{a}_q$ acting on the particle aggregate:

$$\bar{\boldsymbol{P}} = \frac{1}{V} \sum_{q=1}^{Q} \mathbf{a}_q \otimes \boldsymbol{X}_q. \tag{23}$$

The Hill-Mandel micro-heterogeneity condition expresses the equality between the volume average of the virtual work applied at the boundaries of the micro-structure and the virtual work of a macroscopic material point [31]. For a discrete particle system this condition specifies into

$$\bar{\boldsymbol{P}} : \delta\bar{\boldsymbol{F}} = \frac{1}{V} \sum_{q=1}^{Q} \mathbf{a}_q \cdot \delta\mathbf{x}_q. \tag{24}$$

The macroscopic stress $\bar{\boldsymbol{P}}$ given by (23) must satisfy the energy consistency between the two scales. Accordingly, considering definition (15), the following identity holds

$$\bar{\boldsymbol{P}} : \delta\bar{\boldsymbol{F}} = \bar{\boldsymbol{P}} : \frac{1}{V} \sum_{q=1}^{Q} \delta\mathbf{x}_q \otimes \mathbf{A}_q = \frac{1}{V} \sum_{q=1}^{Q} \left(\bar{\boldsymbol{P}} \cdot \mathbf{A}_q\right) \cdot \delta\mathbf{x}_q. \tag{25}$$

Alternatively, by making use of the definition of the macro-scale stress (23), the inner product $\bar{\boldsymbol{P}} : \delta \bar{\boldsymbol{F}}$ can be expanded as

$$\bar{\boldsymbol{P}} : \delta \bar{\boldsymbol{F}} = \frac{1}{V} \sum_{q=1}^{Q} \mathbf{a}_q \otimes \mathbf{X}_q : \delta \bar{\boldsymbol{F}} = \frac{1}{V} \sum_{q=1}^{Q} \mathbf{a}_q \cdot \left( \delta \bar{\boldsymbol{F}} \cdot \mathbf{X}_q \right) = \frac{1}{V} \sum_{q=1}^{Q} \left( \bar{\boldsymbol{P}} \cdot \mathbf{A}_q \right) \cdot \left( \delta \bar{\boldsymbol{F}} \cdot \mathbf{X}_q \right) . \tag{26}$$

Subsequently, reformulating equation (24) as

$$\frac{1}{V} \sum_{q=1}^{Q} \mathbf{a}_q \cdot \delta \mathbf{x}_q - \bar{\boldsymbol{P}} : \delta \bar{\boldsymbol{F}} - \bar{\boldsymbol{P}} : \delta \bar{\boldsymbol{F}} + \bar{\boldsymbol{P}} : \delta \bar{\boldsymbol{F}} = 0 , \tag{27}$$

and substituting equations (25) and (26), leads to

$$\frac{1}{V} \sum_{q=1}^{Q} \left( \mathbf{a}_q - \bar{\boldsymbol{P}} \cdot \mathbf{A}_q \right) \cdot \left( \delta \mathbf{x}_q - \delta \bar{\boldsymbol{F}} \cdot \mathbf{X}_q \right) = 0 . \tag{28}$$

Invoking the micro-scale displacement fluctuations in accordance with relation (14) turns expression (28) finally into

$$\frac{1}{V} \sum_{q=1}^{Q} \left( \mathbf{a}_q - \bar{\boldsymbol{P}} \cdot \mathbf{A}_q \right) \cdot \delta \mathbf{w}_q = 0 . \tag{29}$$

Note that the recast form (29) of the Hill-Mandel condition is satisfied for all three types of boundary conditions introduced above: For the (D) boundary condition, the combination of equations (14) and (17) results in $\delta \mathbf{w}_q = \mathbf{0}$. For the (P) boundary condition, the periodicity of the micro-fluctuations of the boundary displacements $\mathbf{w}_q^+ = \mathbf{w}_q^-$ and the anti-periodicity of the boundary forces $\mathbf{a}_q^+ = -\mathbf{a}_q^-$, following from equations (19) and (20), respectively, make that their products in expression (29) vanish for opposite boundaries. For the (T) boundary condition, relation (21) leads to $\mathbf{a}_q - \bar{\boldsymbol{P}} \cdot \mathbf{A}_q = \mathbf{0}$.

It should be mentioned that the Hill-Mandel condition elaborated above only accounts for the influence of contact forces acting on inner particles, and does not include the effect of contact moments. Although in the frictional contact law the contact moments are absent, see expression (6), for the cohesive contact law they contribute both to the elastic behavior and the strength criterion, see expressions (7) and (8), respectively. The extension of a contact law with a contact moment contribution formally introduces a couple stress in the macroscopic response of the particle aggregate, which is energetically conjugated to the gradient of the overall rotation, see e.g., [32, 33, 34, 35, 36, 37]. These higher-order stress and deformation measures correspond to higher-order natural and essential boundary data [37], which are known to be difficult to measure in experiments, and commonly are (substantially) lower in magnitude than the classical boundary data. For these reasons, and from the fact that the cohesive contact law defined by equations (7) and (8) is used only for one example discussed at the end of this communication, see Section 5.3, an extension of the Hill-Mandel condition with the effect of a contact moments is omitted here, but may be considered as a topic for future research. Correspondingly, for the cohesive contact law a consistency in energy between the micro- and macro- scales of a particle aggregate can only be warranted in an approximate fashion.

In Sections 4 and 5, the results of the DEM analyses will be presented in terms of components of the macro-scale Cauchy stress tensor $\bar{\boldsymbol{\sigma}}$. This stress measure can be derived from the first Piola-Kirchhoff stress $\bar{\boldsymbol{P}}$ computed through (23) by using the common transformation rule:

$$\bar{\boldsymbol{\sigma}} = \frac{1}{\det \left( \bar{\boldsymbol{F}} \right)} \bar{\boldsymbol{P}} \cdot \bar{\boldsymbol{F}}^T . \tag{30}$$

## 3 Numerical implementation of micro-scale boundary conditions

The micro-scale boundary conditions outlined above were implemented by using the open-source discrete element code ESyS-Particle [38, 39]. The numerical algorithms developed for this purpose are described below.

### 3.1 Homogeneous deformation and zero rotation (D)

The homogeneous deformation and zero rotation boundary condition (D) given by equation (17) can be implemented straightforwardly by imposing this condition in an incremental fashion on the boundary particles $\mathcal{P}_q$. After moving the boundary particles in accordance with the incremental update of the deformation $\bar{\boldsymbol{F}}$, dynamic relaxation is applied to reach the equilibrium state of the particle aggregate, during which the displacements imposed on the boundary particles remain fixed. The particle configuration corresponding to the equilibrium state is stored, and the next deformation increment is applied. This process is repeated until the total number of deformation increments $i_{tot}$ is reached. The details of the algorithm are summarized in Table 1.

Table 1: Algorithm for the (D) boundary condition.

## 3.2 Periodic displacement/periodic rotation (P) and uniform force/free rotation (T)

The periodic displacement and periodic rotation boundary condition (P) and the uniform force and free rotation boundary condition (T) were numerically implemented by means of a servo-control algorithm, which uses a feedback principle similar to that of algorithms commonly applied within control theory of dynamic systems [19]. More specifically, the algorithms iteratively correct the boundary particle displacements and rotations from a gradually diminishing discrepancy between the measured and the required values of the boundary condition.

For the periodic displacement and periodic rotation boundary condition (P), the boundary forces and boundary moments should satisfy the anti-periodicity conditions presented in equation (20). Accordingly, the corresponding residuals for the edge particles are

$$\Delta \mathbf{a}_e = \mathbf{a}_e^+ + \mathbf{a}_e^-, \quad \Delta m_e = (\mathbf{m}_e^+ + \mathbf{m}_e^-) \cdot \mathbf{e}_3 \quad \text{for } e = 1..., E/2 \,. \tag{31}$$

Multiplying the residuals by corresponding gain parameters $g_a^p$ and $g_m^p$ results into the following displacement and rotation corrections for the edge particles:

$$\Delta \mathbf{u}_e^+ = \Delta \mathbf{u}_e^- = g_a^p \Delta \mathbf{a}_e, \quad \Delta \theta_e^+ = \Delta \theta_e^- = g_m^p \Delta m_e \quad \text{for } e = 1..., E/2 \,, \tag{32}$$

which are added to the particle locations and rotations from the previous iteration. Note that the four corner particles straightforwardly follow the macroscopic deformation $\bar{\boldsymbol{F}}$, by prescribing their displacements in accordance with equation (17). Hence, for these particles no displacement correction is needed. The rotations of the four corner particles will be updated similarly to (32), using the corrections

$$\Delta \theta_c^+ = \Delta \theta_c^- = g_m^p \Delta m_c \quad \text{with} \quad \Delta m_c = \sum_{c=1}^{4} \boldsymbol{m}_c \cdot \boldsymbol{e}_3 \,. \tag{33}$$

For the uniform force and free rotation boundary condition (T), the boundary forces ensue from the applied macroscopic stress through relation (21), whereby the imposed macroscopic deformation is given by expression (15). The corresponding residuals can be formulated as

$$\Delta \mathbf{a}_q = \bar{\boldsymbol{P}} \cdot \mathbf{A}_q - \mathbf{a}_q, \quad \Delta \bar{\boldsymbol{F}}_q = \sum_{r=1}^{Q} \left[ V \bar{\boldsymbol{F}} \cdot \mathbf{A}_q - (\mathbf{A}_q \cdot \mathbf{A}_r) \, \mathbf{x}_r \right] \quad \text{for } q = 1..., Q \,. \tag{34}$$

The correction for the displacement of the boundary particles is derived by multiplying the force and deformation residuals in (34) by the gain parameters $g_a^t$ and $g_F^t$, respectively, leading to

$$\Delta \mathbf{u}_q = g_a^t \Delta \mathbf{a}_q + g_F^t \Delta \bar{\boldsymbol{F}}_q \quad \text{for } q = 1..., Q \,. \tag{35}$$

When performing numerical simulations, the specific values of the gain parameters $g_a^p$, $g_m^p$, $g_a^t$, $g_F^t$ need to be fine-tuned from accuracy and stability considerations of preliminary numerical benchmark tests.

The corrections for the displacement and rotation of the boundary particles were implemented by means of two different algorithms, which consider or not an initial prediction of the position of the boundary particles based on their positions calculated at the previous loading step. These algorithms are therefore given the labels "with initial displacement prediction" and "without initial displacement prediction". The algorithms are discussed below, and their effect on the computational results will be investigated in Section 4. The specific parts of the algorithms that refer to the periodic displacement and periodic rotation boundary condition will be denoted by the symbol (P), while the symbol (T) indicates the uniform force and free rotation boundary condition. Finally, the residuals defined in expressions (31) and (34), which relate to the particle force, particle moment and macroscopic deformation gradient, are evaluated at each iteration by subjecting their dimensionless form to a convergence check. The dimensionless forms are obtained through, respectively, a normalization by the following parameters:

$$\tilde{a}_k = \frac{M_k R_k}{\Delta t^2}, \qquad \tilde{m}_k = \frac{M_k R_k^2}{\Delta t^2}, \qquad \tilde{F}_k = R_k^3 \,, \tag{36}$$

with $k = c, e, q$ referring to corner, edge, and boundary particles, respectively. In (36), $M_k$ is the mass of particle $k$, $R_k$ is its radius and $\Delta t$ is the time increment used in the dynamic relaxation procedure.

### 3.2.1 Algorithm with initial displacement prediction

The macroscopic deformation is imposed in $i_{tot}$ steps on the boundary particles $\mathcal{P}_q$ via the incrementally updated deformation gradient $\bar{\boldsymbol{F}}$. In correspondence with the algorithm presented in Table 2, in the initialization step, $i = 0$, the boundary particles are moved in accordance with a homogeneous deformation, and for the periodic boundary also a zero rotation, similar to equation (17). Subsequently, the granular assembly is dynamically relaxed to the equilibrium state, keeping the translational and, for the periodic boundary, rotational degrees of freedom of the boundary particles fixed. The iterative loop is entered, and the actual values of the forces and moments of the boundary particles are recorded. For the (P) boundary condition, the corrections for obtaining periodic particle translations and rotations at the boundary are calculated for the corner and edge particles separately, in accordance with relations (31)-(33). For the (T) boundary condition, the boundary moments vanish and the displacement corrections are computed via (34)-(35). The residuals are computed and compared with prescribed tolerances. For the (P) boundary condition, the residual is based on boundary forces and moments. For the (T) boundary condition, two residuals are calculated, which are based on the boundary forces and on the imposed macroscopic deformation. If the norm of the residual(s) is(are) smaller than the tolerance(s) (referred to as $\epsilon_a$ for the force criterion and $\epsilon_F$ for the deformation criterion), the iterative loop is terminated and the next loading step is applied. If the convergence criterion is not satisfied, the corrections are computed again and the residual is iteratively re-examined, until convergence is reached.

After the initialization step is concluded, the responses for subsequent increments, $1 \leq i \leq i_{tot}$, are calculated, see Table 2. For the (P) boundary condition, the corner nodes are moved by straightforwardly imposing the updated macro-scale deformation in accordance with relation (17). For the edge particles, their current position is determined from a prediction based on the particle position in the previous loading step $i - 1$. More specifically, this prediction is a function of the position a particle would have in case of a homogeneous deformation (using the displacement boundary condition (17)), plus the difference, multiplied by an inheritance factor $n_f$, between the final particle position at the previous increment and the position the particle would have at the previous increment under a homogeneous deformation. The inheritance factor lies between 0 and 1, and its optimal value depends on the loading conditions applied and the characteristics of the particle assembly. For the (T) boundary condition, the prediction occurs in an analogous fashion and is applied to all the boundary particles. After the boundary particles are translated in accordance with the predicted values of their positions, the granular assembly is dynamically relaxed to its equilibrium state. Subsequently, the iterative loop is entered, which invokes the previously described correction procedure of the displacements and rotations, in correspondence with the servo-control methodology.

### 3.2.2 Algorithm without initial displacement prediction

Similar to the algorithm with initial displacement prediction, for the algorithm without initial displacement prediction the macroscopic deformation is imposed in $i_{tot}$ steps to the boundary particles $\mathcal{P}_q$. However, as pointed out in Table 3, all increments are now treated in the same fashion. The boundary particles are initially moved in accordance with the updated homogeneous macroscopic deformation $\bar{\boldsymbol{F}}$, similar to expression (17), after which the particle assembly is dynamically relaxed to its equilibrium state. The iterative loop is started, in which the corrections for the displacement and rotation of the boundary particles are calculated based on the servo-control methodology. For the (P) boundary condition, the boundary is partitioned into corner and edge particles, whereby relations (31)-(33) are applied. For the (T) boundary condition, equations (34)-(35) are employed. Subsequently, the particle system is relaxed to the equilibrium state, and the current values of the boundary forces and moments are recorded and used to compute the residuals. If the norms of the residuals are smaller than the corresponding tolerances adopted, the iterative loop is terminated and the next loading step is applied. It can be confirmed that the algorithm without displacement prediction can be obtained as a limit case of the algorithm with initial displacement prediction by setting the inheritance factor equal to zero, $n_f = 0$, whereby the algorithmic structure provided in Table 2 reduces to the more compact and simpler algorithmic structure presented in Table 3.

## 4 Computational results for regular and irregular packings

The algorithms proposed above for the implementation of the micro-scale boundary conditions are tested on a series of DEM simulations on regular, monodisperse and irregular, polydisperse particle packings.

**ALGORITHM WITH INITIAL DISPLACEMENT PREDICTION**

---

**1. Initialization DEM simulation. Increment** $i = 0$
   **1.1** Initialize boundary conditions by applying updated macro-scale deformation homogeneously
      **1.1.A** if (P) $\implies$ $\mathbf{x}_q = \bar{\mathbf{F}}\mathbf{X}_q$ and $\mathbf{Q}_q = \mathbf{I}$ for $q = 1..., Q$
      **1.1.B** if (T) $\implies$ $\mathbf{x}_q = \bar{\mathbf{F}}\mathbf{X}_q$ and $\boldsymbol{m}_q = \mathbf{0}$ for $q = 1..., Q$
   **1.2** Dynamic relaxation. Obtain boundary forces and moments.
   **1.3** Update particle configuration
      **1.3.A** if (P) $\implies$ Partition the boundary into corner $c$ and edge $e$ particles
         Calculate edge particles displacement $\Delta \mathbf{u}_e$ and rotation $\Delta \theta_e$ corrections via (31)-(32)
         and corner particles rotation corrections $\Delta \theta_c$ via (33)
      **1.3.B** if (T) $\implies$ Calculate boundary particles displacement correction $\Delta \mathbf{u}_q$ via (34)-(35)
   **1.4** Dynamic relaxation. Obtain boundary forces and moments.
   **1.5** Calculate residual(s)
      **1.5.A** if (P) $\implies$ $r_a = \sqrt{\sum_{e=1}^{E/2}(\Delta \mathbf{a}_e \cdot \Delta \mathbf{a}_e/\tilde{a}_e^2 + (\Delta m_e/\tilde{m}_e)^2) + (\Delta m_c/\tilde{m}_c)^2}$
      **1.5.B** if (T) $\implies$ $r_a = \sqrt{\sum_{q=1}^{Q} \Delta \mathbf{a}_q \cdot \Delta \mathbf{a}_q/\tilde{a}_q^2}$ and $r_F = \sqrt{\sum_{q=1}^{Q} \Delta \bar{\boldsymbol{F}}_q \cdot \Delta \bar{\boldsymbol{F}}_q/\tilde{F}_q^2}$
   **1.6** Check for convergence: $r_a \leq \epsilon_a$ for (P); $r_a \leq \epsilon_a$ and $r_F \leq \epsilon_F$ for (T)
      **1.6.A** if converged $\implies$ Save current configuration and go to 2
      **1.6.B** if not converged $\implies$ Return to 1.3

---

**2. Subsequent increments** $1 \leq i \leq i_{tot}$
   **2.1** Apply updated boundary conditions
      **2.1.A** if (P) $\implies$
         Impose updated macro-scale deformation on corner nodes: $\mathbf{x}_c = \bar{\boldsymbol{F}}\mathbf{X}_c$ and $\mathbf{Q}_q = \mathbf{I}$
         Prediction of the positions of edge particles:
         $\mathbf{x}_e^i = \mathbf{x}_e^{i,(D)} + n_f \left( \mathbf{x}_e^{i-1} - \mathbf{x}_e^{i-1,(D)} \right)$ for $e = 1...E/2$
         with $\mathbf{x}_e^{(D)} = \bar{\boldsymbol{F}}\mathbf{X}_e$ and the inheritance factor $0 < n_f \leq 1$
      **2.1.B** if (T) $\implies$
         Prediction of the positions of boundary particles:
         $\mathbf{x}_q^i = \mathbf{x}_q^{i,(D)} + n_f \left( \mathbf{x}_q^{i-1} - \mathbf{x}_q^{i-1,(D)} \right)$ for $q = 1..., Q$
         with $\mathbf{x}_q^{(D)} = \bar{\boldsymbol{F}}\mathbf{X}_q$ and the inheritance factor $0 < n_f \leq 1$
   **2.2** Translate particles according to predictions 2.1.A, or 2.1.B
   **2.3** Dynamic relaxation. Obtain boundary forces and moments.
   **2.4** Update particle configuration with displacement and rotation corrections $\Delta \mathbf{u}$ and $\Delta \theta$
      **2.4.A** if (P) $\implies$ Refer to 1.3.A
      **2.4.B** if (T) $\implies$ Refer to 1.3.B
   **2.5** Dynamic relaxation. Obtain boundary forces and moments.
   **2.6** Calculate residual(s)
      **2.6.A** if (P) $\implies$ Refer to 1.5.A
      **2.6.B** if (T) $\implies$ Refer to 1.5.B
   **2.7** Check for convergence: $r_a \leq \epsilon_a$ for (P); $r_a \leq \epsilon_a$ and $r_F \leq \epsilon_F$ for (T)
      **2.7.A** if converged $\implies$ Save current configuration and go to 2 (next increment $i + 1$)
      **2.7.A** if not converged $\implies$ Return to 2.4

---

Table 2: Algorithm for the (P) and (T) boundary conditions *with* initial displacement prediction.

## 4.1 Regular monodisperse packing

In this section the responses of three different regular, monodisperse particle packings are considered, which consist of circular particles of radius $R = 1.02$ mm, where the centroids of two particles in contact initially are at a distance of $2.0$ mm. The initial volumes of the packings are $V = [64, 324, 784]$ mm$^2$, which are calculated from the locations of the centroids of the four corner particles. The number of particles of the three prackings are equal to $n_p = [25, 100, 225]$. The particle volume fraction related to initial volume occupied by the inner particles is $\mathfrak{v} = 0.785$. The corresponding coordination number, which reflects the average number of contacts of the inner particles, equals 4.

## ALGORITHM WITHOUT INITIAL DISPLACEMENT PREDICTION

**1. DEM simulation. Increments** $0 \leq i \leq i_{tot}$
    **1.1** Initialize boundary conditions by applying updated macro-scale deformation homogeneously
        **1.1.A** if (P) $\implies \mathbf{x}_q = \bar{\mathbf{F}} \mathbf{X}_q$    and    $\mathbf{Q}_q = \mathbf{I}$    for $q = 1..., Q$
        **1.1.B** if (T) $\implies \mathbf{x}_q = \bar{\mathbf{F}} \mathbf{X}_q$    and    $\boldsymbol{m}_q = \mathbf{0}$    for $q = 1..., Q$
    **1.2** Dynamic relaxation. Obtain boundary forces and moments.
    **1.3** Update particle configuration
        **1.3.A** if (P) $\implies$ Partition the boundary into corner *c* and edge *e* particles
            Calculate edge particles displacement $\Delta \mathbf{u}_e$ and rotation $\Delta \theta_e$ corrections via (31)-(32)
            and corner particles rotation corrections $\Delta \theta_c$ via (33)
        **1.3.B** if (T) $\implies$ Calculate boundary particles displacement correction $\Delta \mathbf{u}_q$ via (34)-(35)
    **1.4** Dynamic relaxation. Obtain boundary forces and moments.
    **1.5** Calculate residual(s)
        **1.5.A** if (P) $\implies r_a = \sqrt{\sum_{e=1}^{E/2}(\Delta \mathbf{a}_e \cdot \Delta \mathbf{a}_e / \tilde{a}_e^2 + (\Delta m_e / \tilde{m}_e)^2) + (\Delta m_c / \tilde{m}_c)^2}$
        **1.5.B** if (T) $\implies r_a = \sqrt{\sum_{q=1}^{Q} \Delta \mathbf{a}_q \cdot \Delta \mathbf{a}_q / \tilde{a}_q^2}$ and $r_F = \sqrt{\sum_{q=1}^{Q} \Delta \bar{\boldsymbol{F}}_q \cdot \Delta \bar{\boldsymbol{F}}_q / \tilde{F}_q^2}$
    **1.6** Check for convergence: $r_a \leq \epsilon_a$ for (P); $r_a \leq \epsilon_a$ and $r_F \leq \epsilon_F$ for (T)
        **1.6.A** if converged $\implies$ Save current configuration and go to 1 (next increment $i+1$)
        **1.6.B** if not converged $\implies$ Return to 1.3

Table 3: Algorithm for the (P) and (T) boundary conditions *without* initial displacement prediction.

The particles obey a frictional contact law, in correspondence with relation (6). Assuming relatively soft particles, the normal and tangential stiffnesses are chosen as $k_n = 10^4$ N/m and $k_s = 2 \cdot 10^3$ N/m, and the friction coefficient equals $\mu = 0.4$. The density of the particles is $\rho = 2 \cdot 10^3$ kg/m$^2$. The translational and rotational damping factors used in the dynamic relaxation procedure are $\alpha = \beta = 0.7$. The packings are subjected to a combined biaxial compression-true shear deformation

$$\bar{\boldsymbol{F}} = \mathbf{I} + \bar{F}_{11} \, \boldsymbol{e}_1 \otimes \boldsymbol{e}_1 + \bar{F}_{12} \, \boldsymbol{e}_1 \otimes \boldsymbol{e}_2 + \bar{F}_{21} \, \boldsymbol{e}_2 \otimes \boldsymbol{e}_1 + \bar{F}_{22} \, \boldsymbol{e}_2 \otimes \boldsymbol{e}_2 \,, \tag{37}$$

with $\bar{F}_{11} = \bar{F}_{22} = -0.03$ and $\bar{F}_{12} = \bar{F}_{21} = -0.3$, which is applied in $i_{tot} = 300$ loading steps. For reaching the equilibrium state at each loading step, the particle system is subjected to dynamic relaxation steps of constant time increments $\Delta t = 10^{-6}$s. The gain parameters (in dimensionless form) used for the correct application of the boundary conditions are: for (P) $g_a^p M/\Delta t^2 = 1 \cdot 10^2$ and $g_m^p M R^2 / \Delta t^2 = 2 \cdot 10^2$; for (T) $g_a^t M / \Delta t^2 = 1 \cdot 10^2$ and $g_F^t R^2 = 2 \cdot 10^{-5}$, with $M = \rho \pi R^2$ representing the mass of the particles. The force and deformation tolerances are taken as $\epsilon_a = 10^{-6}$ and $\epsilon_F = 10^{-2}$, respectively. For the dynamic relaxation process, a value of $10^{-3}$ is adopted for tol$_E$, whereby equation (11) must be minimally satisfied for a pre-defined, continuous period of $20\Delta t$, in order to ensure a rigorous dynamic relaxation to the equilibrium state. An overview of the model parameters is given in Table 4.

### 4.1.1 Responses for algorithms with and without initial displacement prediction

In order to investigate the performance of the two algorithms presented in Tables 2 and 3, the packing of 25 particles is considered first. The stress responses under the combined biaxial compression-true shear loading were computed with Eq.(30), and plotted as a function of the applied macroscopic shear deformation $\bar{F}_{12}$. Figures 2(a) and (b) show the results for the algorithms with (solid line) and without (dot-dashed line) an initial displacement prediction for the (P) and (T) boundary conditions, respectively. The normal and shear components of the Cauchy stress are normalized as $\tilde{\sigma}_{11} = \bar{\sigma}_{11} R / k_n$ and $\tilde{\sigma}_{12} = \bar{\sigma}_{12} R / k_n$, respectively, where $\bar{\sigma}_{11}$ and $\bar{\sigma}_{12}$ are the macroscopic normal and shear Cauchy stresses of the particle aggregate. For the periodic boundary conditions (P), Figure 3 illustrates the packing structures at specific macroscopic shear deformations $\bar{F}_{12} = -0.01$ (a), $\bar{F}_{12} = -0.113$ (b), and $\bar{F}_{12} = -0.28$ (c). The red lines plotted in the deformed particle aggregates indicate the network of normal contact forces between the particles.

For the algorithm with initial displacement prediction, the local minimum of the normal stress $\tilde{\sigma}_{11}$ near $\bar{F}_{12} = -0.113$, as shown in Figure 2(a) for the periodic boundary conditions (P), can be ascribed to a joint localized sliding of all the boundary particles, see Figure 3(b), top. This localization mechanism does not arise for the algorithm without initial displacement prediction, which furnishes a shear response that is much more homogeneous, see Figure 3(b), bottom. It may be therefore concluded that the response of the packings is rather sensitive to bifurcations in the equilibrium path followed, which here become evident due to the relatively low number of particles present in the packing. Under continuing deformation towards $\bar{F}_{12} = -0.28$, the inner particles of the aggregate also develop substantial sliding, such that for both algorithms the particle structure gradually reaches its densest packing structure, at which the deformation as well as the normal contact force network become strongly homogeneous. Both for the normal and shear stress components the responses computed by the two algorithms at this stage have coalesced, and steadily grow under further increasing deformation.

| Parameter | Value | Unit |
|---|---|---|
| Elastic normal stiffness $k_n$ | $1 \cdot 10^4$ | N/m |
| Elastic tangential stiffness $k_s$ | $2 \cdot 10^3$ | N/m |
| Friction coefficient $\mu$ | 0.4 | - |
| Density $\rho$ | $2 \cdot 10^3$ | kg/m$^2$ |
| Translational damping $\alpha$ | 0.7 | - |
| Rotational damping $\beta$ | 0.7 | - |
| Time increment $\Delta t$ | $10^{-6}$ | s |
| Tolerance force (P), (T) $\epsilon_a$ | $10^{-6}$ | - |
| Tolerance deformation (T) $\epsilon_F$ | $10^{-2}$ | - |
| Gain force (P) $g_a^p M/\Delta t^2$ | $1 \cdot 10^2$ | - |
| Gain moment (P) $g_m^p MR^2/\Delta t^2$ | $2 \cdot 10^2$ | - |
| Gain force (T) $g_a^t M/\Delta t^2$ | $1 \cdot 10^2$ | - |
| Gain deformation (T) $g_F^t R^2$ | $2 \cdot 10^{-5}$ | - |
| Tolerance dynamic relaxation $\text{tol}_E$ | $10^{-3}$ | - |

Table 4: Physical and algorithmic model parameters.

A similar trend can be observed for the normal and shear stress responses of the particle aggregates with the (T) boundary condition, see Figure 2(b). The discrepancies in the responses computed by the two algorithms appears to be less than for the (P) boundary condition. Since the computational robustness and efficiency of the two algorithms is comparable in the present simulations, no clear distinction can be made in terms of their overall performance. Hence, as an arbitrary choice, the forthcoming DEM results are computed with the algorithm without initial displacement prediction.

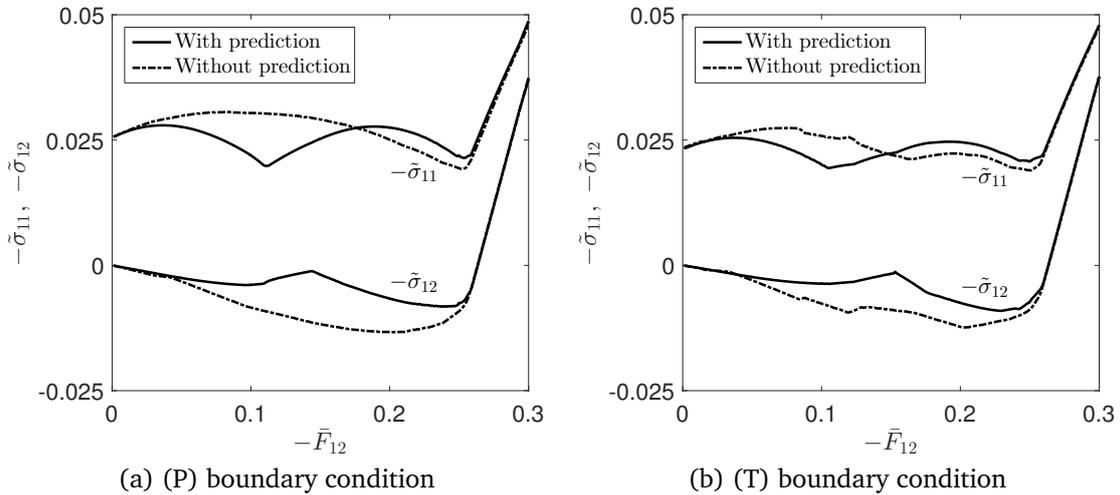

(a) (P) boundary condition  (b) (T) boundary condition

Figure 2: Normalized macroscopic Cauchy stresses $-\tilde{\sigma}_{11}$ and $-\tilde{\sigma}_{12}$ versus the shear deformation $-\bar{F}_{12}$ for (a) periodic displacement/periodic rotation boundary condition (P) and (b) uniform force/free rotation boundary condition (T). The responses relate to a regular monodisperse packing of 25 particles, and were computed by the algorithms with (solid line) and without (dot-dashed line) initial displacement predictions.

### 4.1.2 Responses for the (D), (P) and (T) boundary conditions

The influence of the choice of the boundary condition on the overall packing response is illustrated in Figure 4(a). The (normalized) normal stress $\tilde{\sigma}_{11}$ is shown as a function of the applied shear deformation $\bar{F}_{12}$ for the displacement/zero rotation boundary condition (D) with a solid line, for the periodic displacement/periodic rotation boundary condition (P) with a dot-dashed line, and for the uniform force/free rotation boundary condition (T) with a dashed line. The stress response computed for the (P) boundary condition is bounded by the stiffer and softer responses measured for the (D) and (T) boundary conditions, respectively, a result that is in agreement with the numerical studies performed in [16]. It may be observed that the initial stress value corresponding to the (T) boundary condition is somewhat smaller than the value computed for the other two boundary conditions. This is, because after the sample preparation procedure was finished, the boundary forces generated by a 4% particle overlap do not exactly satisfy equation (21), and therefore in the first loading increment are slightly relaxed by the algorithm in order to meet this condition.

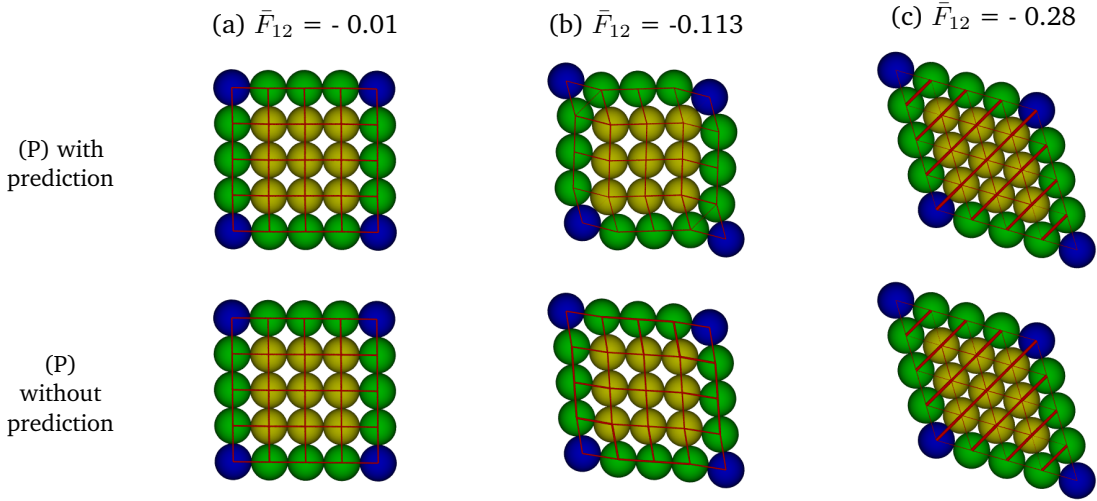

Figure 3: Deformed configurations of a regular packing of 25 particles with (P) boundary conditions evaluated at three different deformation states: (a) $\bar{F}_{12} = -0.01$, (b) $\bar{F}_{12} = -0.113$, (c) $\bar{F}_{12} = -0.28$. The particle configurations were computed with the algorithms with (top) and without (bottom) initial displacement prediction. The red lines indicate the normal contact force network of the particles.

Consider the average normalized particle overlap $\Delta \bar{u}_n$, defined as

$$\Delta \bar{u}_n = \frac{1}{N^c} \sum_{c=1}^{N^c} \frac{\Delta u_n^c}{\bar{R}^c}, \qquad (38)$$

where $N^c$ is total number of particle contacts, $\bar{R}^c$ is the average radius at contact $c$ and $\Delta u_n^c$ is the particle overlap at contact $c$. Figure 4(b) depicts $\Delta \bar{u}_n$ as a function of the applied shear deformation $-\bar{F}_{12}$. Observe that the trend for the average particle overlap is similar to that for the macroscopic normal stress in Figure 4(a). This can be explained as follows: The macroscopic stress is represented by the volume average of all contact forces generated within the granular microstructure. Since the assumed normal contact stiffness is considerably larger than the shear contact stiffness, $k_n >> k_s$, see Table 4, the normal contact forces $f_n^c$, which are proportional to contact overlaps $\Delta u_n^c$, see expression (6), provide the main contribution to the macroscopic stress response.

### 4.1.3 Responses for different sample sizes

The effect of the sample size on the macroscopic stress response is considered by plotting the computational results for packings composed of 25, 100 and 225 particles for the (D), (P) and (T) boundary conditions in Figures 5(a), (b) and (c), respectively. Generally, for a larger sample the normal stress $\tilde{\sigma}_{11}$ decreases. The (D) and (P) boundary conditions show a close resemblance in the responses for 100 and 225 particles, from which it may be concluded that for a sample of about 225 particles the stress response has more or less converged. Conversely, for the (T) boundary condition the stress for a sample of 225 particles shows a substantial relative drop in value up to a deformation of $\bar{F}_{12} \approx 0.20$. This softening behavior appears to be governed by strongly localized deformations emerging at the boundaries of the particle system, a phenomenon that also has been reported for continuum homogenization methods equipped with this relatively soft boundary condition, see [40].

## 4.2 Irregular polydisperse packing

The irregular polydisperse packings analyzed in this section are composed of circular particles, with the particle radii arbitrarily taken from a uniform size distribution with polydispersity $R_{max}/R_{min} = 2$, where $R_{min} = 0.67$

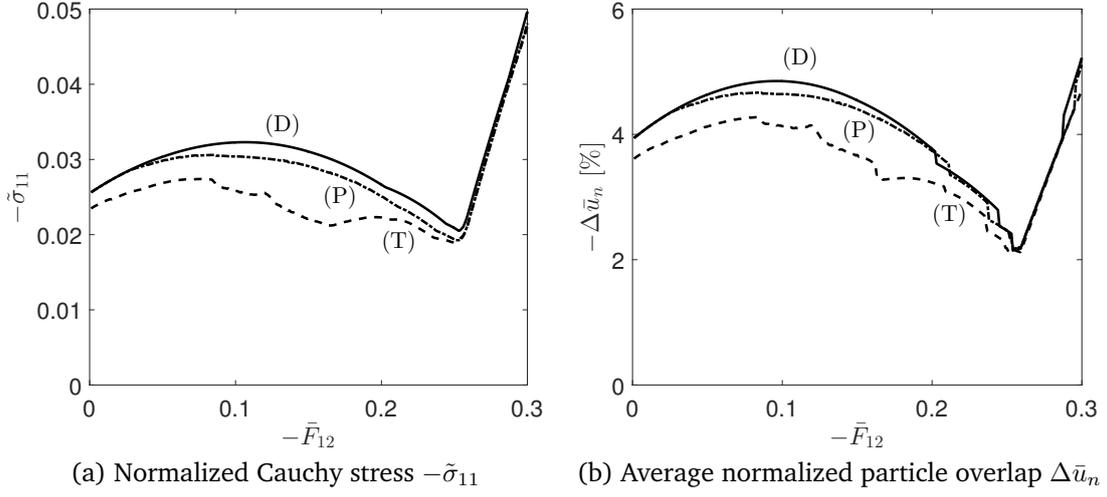

(a) Normalized Cauchy stress $-\tilde{\sigma}_{11}$

(b) Average normalized particle overlap $\Delta \bar{u}_n$

Figure 4: (a) Normalized homogenized Cauchy stress $-\tilde{\sigma}_{11}$ and (b) average particle overlap $\Delta \bar{u}_n$ versus the shear deformation $-\bar{F}_{12}$ for the three types of boundary conditions (D),(P) and (T).

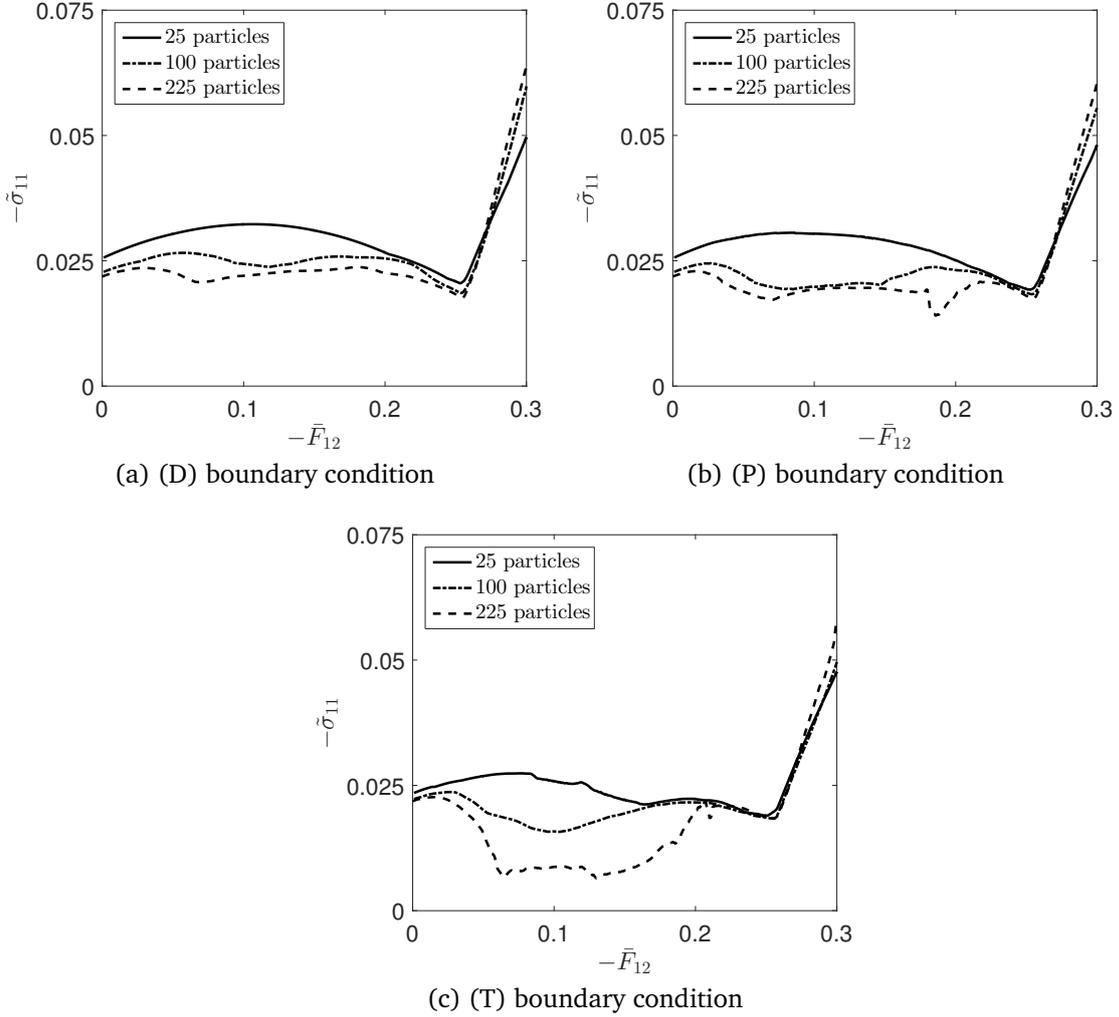

(a) (D) boundary condition

(b) (P) boundary condition

(c) (T) boundary condition

Figure 5: Normalized macroscopic Cauchy stress $-\tilde{\sigma}_{11}$ versus the shear deformation $-\bar{F}_{12}$ for three different sample sizes of 25, 100 and 225 particles, for the (a) (D) boundary condition, (b) (P) boundary condition, and (c) (T) boundary condition.

mm. A collision-driven molecular dynamics code described in [41] is used to randomly generate irregular packings with the initial number of particles equal to $n_p^0 = [25, 100, 200, 400, 600]$, as shown in Figure 6(a). Subsequently, these packings are reconstructed into geometrically periodic packings by copying each of the boundary particles intersecting with the edges of the square particle volume to corresponding positions at the opposite boundaries. This results into the packing geometries shown in Figure 6(b), with the final particle numbers as $n_p = [37, 120, 228, 444, 650]$. The initial volumes of the particle aggregates are equal to $V = [100, 400, 818, 1462, 2156]$ mm$^2$, respectively. The rose diagrams of the particle assemblies are sketched in Figure 6(c), clearly indicating that the packings become more isotropic when the particle number increases. The particle volume fraction of the packings varies in the range $\mathfrak{v} \in [0.833, 0.850]$, where the smallest and highest values correspond to the packings with the smallest and highest number of particles, respectively. The corresponding coordination numbers lie in between 2.97 and 3.47.

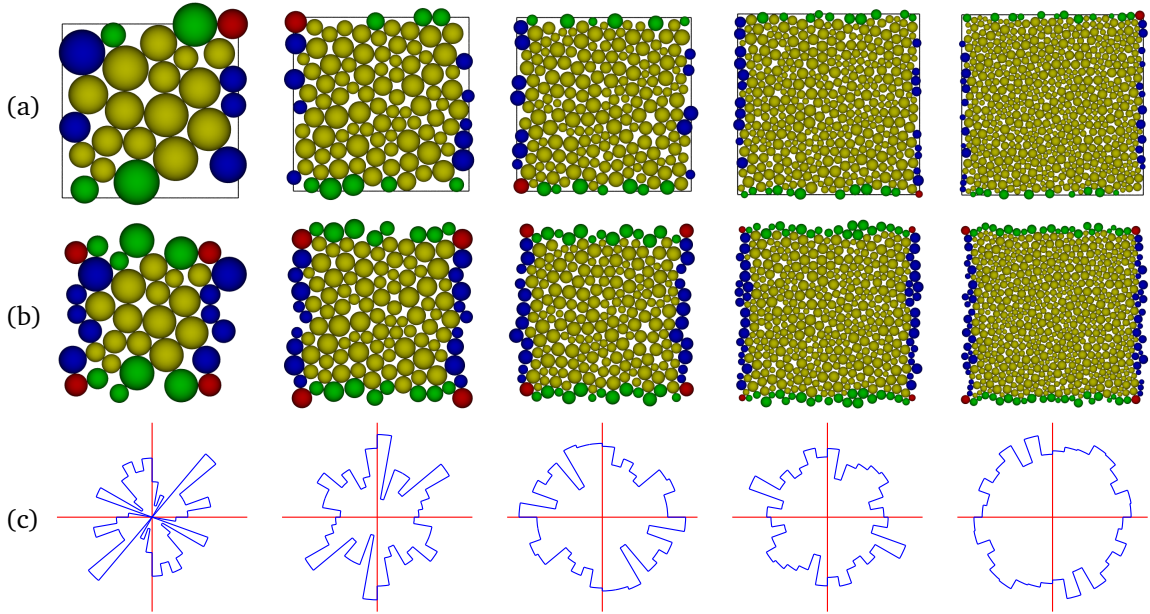

Figure 6: Characteristics of the five different irregular polydisperse packings studied: (a) Initial packings generated by a collision-driven molecular dynamics code [41], (b) geometrically periodic packings with the number of particles equal to $n_p = [37, 120, 228, 444, 650]$, and (c) the rose diagrams.

The particle packings are subjected to a simple shear macroscopic deformation

$$\bar{\boldsymbol{F}} = \mathbf{I} + \bar{F}_{12}\, \boldsymbol{e}_1 \otimes \boldsymbol{e}_2\,, \qquad (39)$$

with $\bar{F}_{12} = 0.5$, which is applied in $i_{tot} = 100$ loading steps. For the irregular packings the same physical and algorithmic parameters are used as for the regular packings, see Table 4, except for the tolerance $\epsilon_F = 10^{-1}$ and the two gain values for the (T) boundary condition, which here relate to $g_a^t M_i/\Delta t^2 = 5$ and $g_F^t R_i^2 = 2 \cdot 10^{-6}$, with $M_i = \rho \pi R_i^2$ being the mass of particle $i$ and $R_i$ its radius. Note that for an irregular polydisperse packing the specific gain values depend on the characteristics of the actual particle $i$.

### 4.2.1 Responses for the (D), (P) and (T) boundary conditions

The response of a packing with 228 particles is considered first. The normalized macroscopic stresses $\tilde{\sigma}_{11} = \bar{\sigma}_{11} \bar{R}/k_n$ and $\tilde{\sigma}_{22} = \bar{\sigma}_{22} \bar{R}/k_n$, with the average radius as $\bar{R} = \sum_{i=1}^{P+Q} R_i/(P+Q)$, are shown in Figure 7 as a function of the applied macroscopic shear deformation $\bar{F}_{12}$. The solid, dot-dashed and dashed lines refer to the (D), (P) and (T) boundary conditions, respectively. Since for the packing of 228 particles the particle structure is rather isotropic, see Figure 6(c), it can be confirmed that the responses for the two normal stresses $\tilde{\sigma}_{11}$ and $\tilde{\sigma}_{22}$ indeed are similar. As for the regular monodisperse packing, the (D) and (T) boundary conditions provide the upper (stiffest) and lower (softest) bounds for the particle system response, and thereby encapsulate the response calculated for the (P) boundary condition. Although not depicted here, the responses for the normalized shear stress $\tilde{\sigma}_{12} = \bar{\sigma}_{12} \bar{R}/k_n$ under the (D), (P) and (T) boundary conditions follow similar trends as observed for the normal stresses $\tilde{\sigma}_{11}$ and $\tilde{\sigma}_{22}$, with the magnitude of the shear stress being about one third of that of the normal stresses.

Figure 8 shows the deformed structure of the granular aggregates for the three types of boundary conditions at four different deformation levels, namely (a) $\bar{F}_{12} = 0.05$, (b) $\bar{F}_{12} = 0.1$, (c) $\bar{F}_{12} = 0.3$ and (d) $\bar{F}_{12} = 0.5$. The local distribution of particles develops differently for the three boundary conditions, leading to differences in the network of normal contact forces represented by the red lines: The (D) and (T) boundary condition experience the highest and lowest contact forces, respectively, as indicated by the relatively thick and thin red lines. Obviously, this is in correspondence with the largest and smallest stress levels for the (D) and (T) boundary conditions, as depicted in Figure 7.

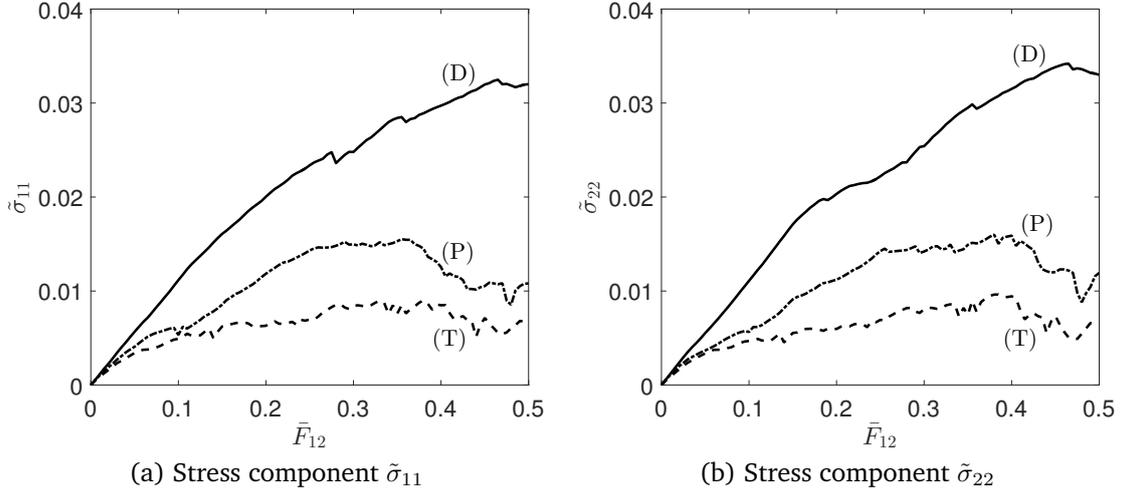

(a) Stress component $\tilde{\sigma}_{11}$

(b) Stress component $\tilde{\sigma}_{22}$

Figure 7: Normalized homogenized Cauchy stresses versus the shear deformation $\bar{F}_{12}$ for packings of 228 particles, subjected to the boundary conditions (D), (P) and (T): (a) Stress component $\tilde{\sigma}_{11}$, (b) stress component $\tilde{\sigma}_{22}$.

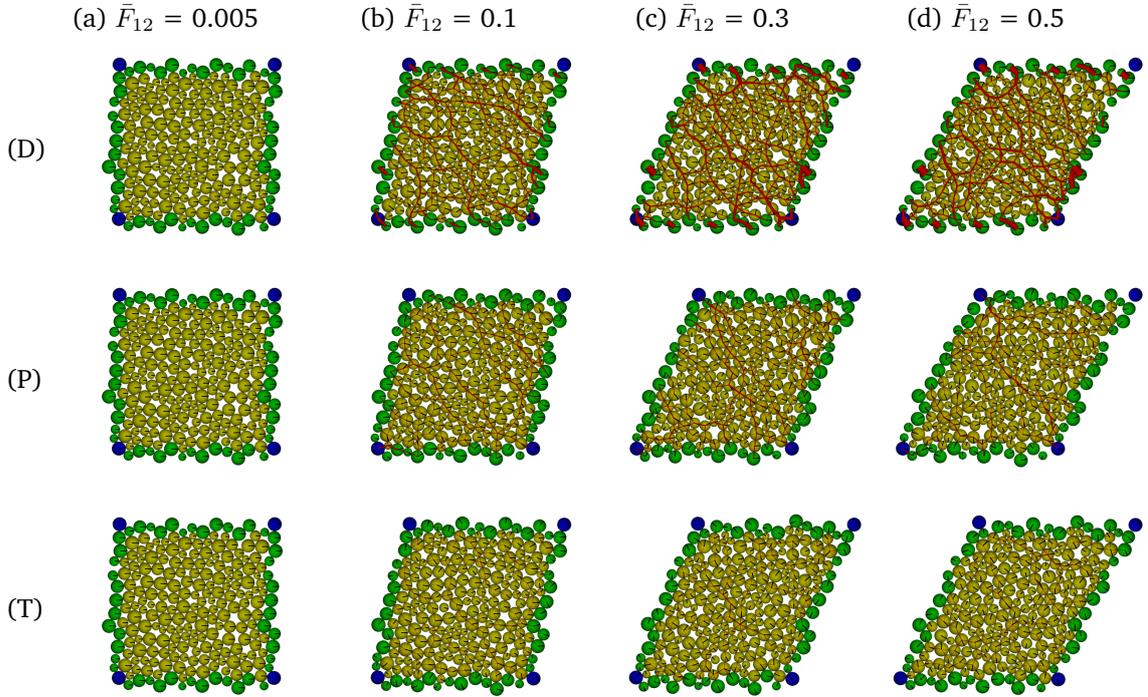

Figure 8: Deformed configurations of an irregular packing of 228 particles evaluated at four different deformation states: (a) $\bar{F}_{12} = 0.05$, (b) $\bar{F}_{12} = 0.1$, (c) $\bar{F}_{12} = 0.3$, and (d) $\bar{F}_{12} = 0.5$, for the (D), (P) and (T) boundary conditions. The red lines indicate the normal contact force network of the particles.

Figure 9 illustrates the average normalized particle overlap $\Delta \bar{u}_n$, defined by relation (38), and the average particle rotation

$$\bar{\theta} = \frac{\sum_{i=1}^{P+Q} |\theta_i|}{P+Q}, \tag{40}$$

both as a function of the applied macroscopic shear deformation $\bar{F}_{12}$. As for the regular monodisperse packings, the average normalized particle overlap is the largest for the (D) boundary condition and the smallest for the (T) boundary condition, and shows a similar evolution as observed for the normal stresses, see Figure 7. As expected, the average rotation shows the opposite trend, being the largest for the soft (T) boundary condition and the smallest for the stiff (D) boundary condition.

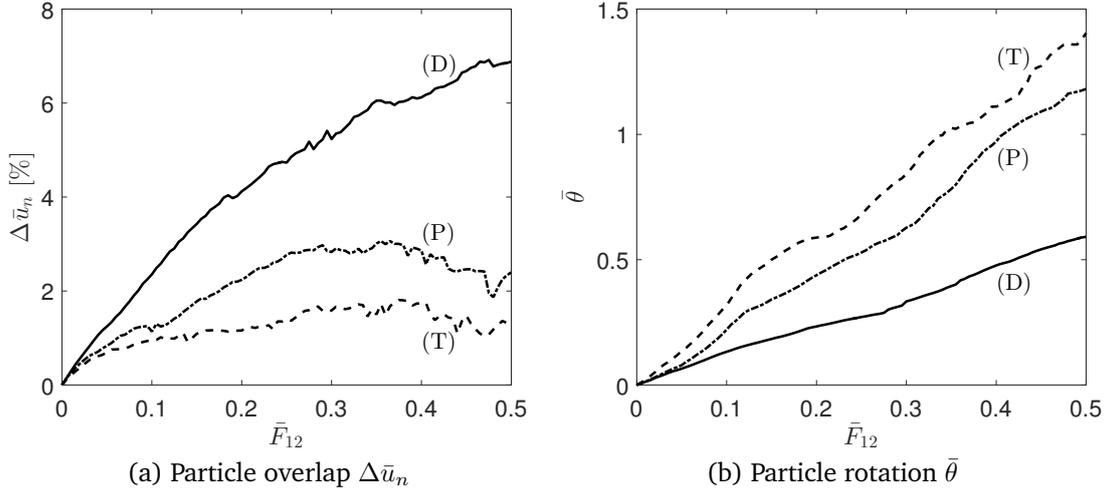

(a) Particle overlap $\Delta \bar{u}_n$

(b) Particle rotation $\bar{\theta}$

Figure 9: (a) Average normalized particle overlap $\Delta \bar{u}_n$ and (b) average particle rotation $\bar{\theta}$ versus the shear deformation $\bar{F}_{12}$ for a packing of 228 particles subjected to the (D), (P) and (T) boundary conditions.

#### 4.2.2 Convergence behavior of macroscopic response under increasing sample size

The convergence behavior of the apparent macroscopic response of the particle aggregate towards its effective response under increasing sample size is studied by subjecting the five microstructures depicted in Figure 6(b) to a simple shear deformation given by (39). In convergence studies, this type of loading condition occasionally is characterized as "critical", because of a relatively slow convergence behavior towards a representative volume element (RVE). The convergence behavior is evaluated here by means of the $L_2$-norm of the normalized, homogenized Cauchy stress tensor $\tilde{\boldsymbol{\sigma}}$, integrated along the entire deformation path

$$\|\tilde{\boldsymbol{\sigma}}\|_{L_2} = \left( \sum_{ij=11,22,12,21} \int_{\bar{F}_{12}=0}^{\bar{F}_{12}=0.5} \tilde{\sigma}_{ij}^2 \, \mathrm{d}\bar{F}_{12} \right)^{1/2}. \tag{41}$$

Figure 10 illustrates the stress norm $\|\tilde{\boldsymbol{\sigma}}\|_{L_2}$ as a function of the sample size, expressed in terms of the number of particles. It can be observed that for the stiff (D) and soft (T) boundary conditions the stress norm, respectively, decreases and increases with increasing sample size, while for the periodic (P) boundary condition it remains approximately constant. These trends are typical for a change in apparent properties under increasing sample size, see e.g., [21]. However, for arriving at an RVE the curves for the (D) (P) and (T) boundary conditions must coincide [31], which indeed is not the case for the largest sample of 650 particles. As already indicated above, the minimum size of the RVE depends on the type of loading condition applied, which is known to be relatively large under a macroscopic shear deformation. From the approximately constant stress value observed in Figure 10 for the (P) boundary condition, it may be expected that the stress response of the RVE will be close to $\|\tilde{\boldsymbol{\sigma}}\|_{L_2} \approx 0.011$. Hence, in multi-scale simulations on granular materials, the computational costs can be kept manageable by adopting the (P) boundary condition for a micro-structural sample of relatively small size, for which the homogenized response is similar to that of the minimal RVE with a much larger size.

## 5 Mixed boundary conditions

In this section the formulation and numerical implementation of *mixed (D)-(P)-(T) boundary conditions* is presented. The homogenization framework proposed satisfies the Hill-Mandel micro-heterogeneity condition, and thus can be used for i) a consistent derivation of macro-scale constitutive relations from standard material tests on particle aggregates subjected to any combination of (D)-, (P)- and/or (T)-type boundary conditions, and ii) the efficient computation of the homogenized response of large-scale particle aggregates characterized by a spatial periodicity in one or two directions, i.e., granular layers exposed to uniform (D) and/or (T) boundary conditions at their top and bottom surfaces. To the best of the authors' knowledge, the formulation presented is novel in the field of granular materials.

### 5.1 Formulation

For the formulation of the mixed boundary conditions, the basic particle configuration sketched in Figure 1 is considered, with the boundary being split up into the top part $\partial V_t$, the bottom part $\partial V_b$, the left part $\partial V_l$ and the right part $\partial V_r$. It is emphasized that the main concepts of the mixed formulation are general, and can be applied to arbitrary boundary value problems. The concepts are elaborated here for the specific case of an infinite horizontal layer of particles loaded by a constant vertical pressure, $\bar{P}_{22} = \bar{P}_{22}^*$, and subsequently subjected to a shear deformation $\bar{F}_{12}$ in horizontal direction. The reason for choosing this boundary value problem is that it includes all the three (D), (P) and (T) boundary conditions discussed previously, with their combinations entering the formulation both at *separate* and *identical* parts of the layer boundary. This allows for highlighting the characteristics of the mixed formulation in full detail. Accordingly, the macroscopic deformation of the particle aggregate is imposed via a *combined* (D)-(T) condition

$$x_{q,1} = \bar{F}_{12} X_{q,2} + X_{q,1} \quad \text{and} \quad a_{q,2} = \bar{P}_{21} A_{q,1} + \bar{P}_{22}^* A_{q,2}, \tag{42}$$

in which the macroscopic shear stress $\bar{P}_{21}$ is measured from the response of the particle assembly. Note that $(42)_1$ implicitly accounts for the condition

$$\bar{F}_{11} = 1. \tag{43}$$

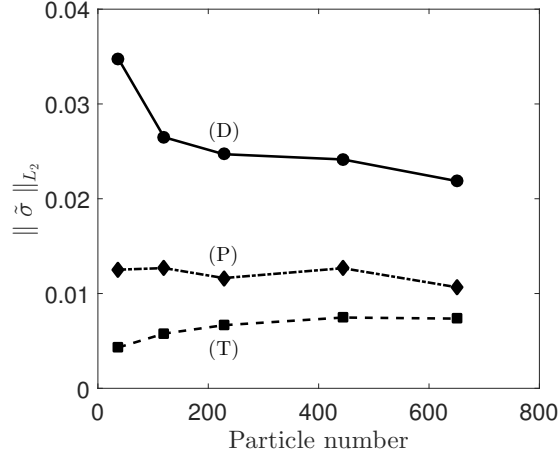

Figure 10: Stress norm $\|\tilde{\sigma}\|_{L_2}$ versus particle number for irregular polydisperse particle packings subjected to (D), (P) and (T) boundary conditions, in accordance with a macroscopic simple shear deformation $\bar{F}_{12}$.

Furthermore, the first contribution in the right-hand side of $(42)_2$ typically is relatively small, since most particles $q$ at the top boundary $\partial V_t$ are characterized by $A_{q,1} << A_{q,2}$, with $A_{q,1}$ vanishing for the specific case of an ideally horizontal boundary composed of identical particles. Since the shear deformation $\bar{F}_{12}$ is imposed *after* the application of the vertical stress $\bar{P}_{22}^*$, in $(42)_1$ the reference positions $\boldsymbol{X}_q$ of the boundary particles relate to the particle configuration obtained after the vertical stress has been applied. In summary, the boundary conditions for the particle aggregate are specified as follows:

- For the particles that are part of the bottom boundary, $q \in \partial V_b$, the homogeneous deformation and zero rotation boundary condition (D) is applied in accordance with expression (17). The vertical boundary displacements are constrained to construct a rigid support for the layer, and the horizontal boundary displacements follow the shear deformation given by expression $(42)_1$.

- For the particles that are part of the left and right boundaries, $q \in \partial V_l \cup \partial V_r$, the periodic displacement and periodic rotation boundary condition (P) is applied, as given by expression (19). This boundary condition reflects the horizontal confinement of the particles within the infinite layer.

- For the particles that are part of the top boundary, $q \in \partial V_t$, free rotations are assumed, in correspondence with the (T) boundary condition[1]. For the description of the particle displacements, the boundary is split up along the two orthonormal directions $\mathbf{e}_1$ and $\mathbf{e}_2$ indicated in Figure 1. Along the $\mathbf{e}_1$-direction, the (D) boundary condition $(42)_1$ is applied to simulate the horizontal macroscopic shear deformation. Along the $\mathbf{e}_2$-direction, a constant macroscopic pressure $\bar{P}_{22}^*$ is imposed via the (T) boundary condition $(42)_2$, for which the corresponding components of the macroscopic deformation gradient, $\bar{F}_{21}$ and $\bar{F}_{22}$, in accordance with the general form (15), turn into

$$\bar{F}_{21} = \frac{1}{V} \sum_{q=1}^{Q} x_{q,2} \otimes A_{q,1} \quad \text{and} \quad \bar{F}_{22} = \frac{1}{V} \sum_{q=1}^{Q} x_{q,2} \otimes A_{q,2}. \tag{44}$$

---

[1] Since the top boundary is subjected to a mixed (D)-(T) boundary condition, instead of leaving the particle rotations free at the boundary, i.e., a (T)-type condition, the particle rotations could have been equally well taken as fully constrained, i.e., a (D)-type condition. For relatively large samples the effect of this choice on the homogenized response of the particle aggregate is expected to be minor.

Note that the two deformation components above should be considered as a computational *result* obtained by prescribing the stress component $\bar{P}_{22}^*$.

The macroscopic deformation gradient $\bar{\boldsymbol{F}}$, which is followed by the four corner nodes of the sample, now is fully specified through its "(D)-type components" provided by $(42)_1$ and (43), and its "(T)-type components" given by $(44)_{1,2}$. The corresponding macroscopic Piola-Kirchhoff stress tensor is defined by equation (23). For the adopted mixed-boundary conditions it will now be demonstrated that this stress definition satisfies the recast Hill-Mandel condition given by expression (29), i.e., the energy consistency between the macro- and micro scales. Accordingly, relation (29) is first split up with respect to the different boundary parts considered above, i.e.,

$$\sum_{q \in \partial V_b} \left(\mathbf{a}_q - \bar{\boldsymbol{P}} \cdot \mathbf{A}_q\right) \cdot \delta\mathbf{w}_q + \sum_{q \in \partial V_l \cup \partial V_r} \left(\mathbf{a}_q - \bar{\boldsymbol{P}} \cdot \mathbf{A}_q\right) \cdot \delta\mathbf{w}_q + \sum_{q \in \partial V_t} \left(\mathbf{a}_q - \bar{\boldsymbol{P}} \cdot \mathbf{A}_q\right) \cdot \delta\mathbf{w}_q = 0. \qquad (45)$$

For the bottom boundary $\partial V_b$, boundary condition (D) holds, which, by comparing equations (17) and (14), lets the micro-fluctuations of the displacements vanish, $\mathbf{w}_q = \mathbf{0}$. Hence, the first term in (45) is equal to zero. At the left and right boundaries $q \in \partial V_l \cup \partial V_r$, the (P) boundary condition is imposed, for which the micro-fluctuations of the displacements are periodic, $\mathbf{w}_q^l = \mathbf{w}_q^r$, see (14) and (19). Together with the anti-periodicity of the boundary forces $\mathbf{a}_q^l + \mathbf{a}_q^r = \mathbf{0}$, see (20), the second term in (45) vanishes. Finally, for the top boundary $\partial V_t$, the last term in (45) may be further developed as

$$\sum_{q \in \partial V_t} \left[\left(a_{q,1} - \bar{P}_{1j} A_{q,j}\right) \delta w_{q,1} + \left(a_{q,2} - \bar{P}_{2j} A_{q,j}\right) \delta w_{q,2}\right] = 0. \qquad (46)$$

Along the $\mathbf{e}_1$-direction, the micro-fluctuations of the boundary particle displacements vanish, $w_{q,1} = 0$, in correspondence with equation $(42)_1$, by which the first term in (46) becomes zero. Along the $\mathbf{e}_2$-direction, the boundary forces are uniform, $a_{q,2} - \bar{P}_{2j} A_{q,j} = 0$, see equation $(42)_2$, so that the second term in (46) becomes zero. With this result, the Hill-Mandel condition (45) is proven to be satisfied for the mixed boundary conditions.

## 5.2 Numerical implementation

The numerical algorithm for the implementation of the mixed boundary conditions is outlined in Table 5, and is based on a combination of the algorithms presented in Section 3 for the (D), (P) and (T) boundary conditions, without an initial displacement prediction.

During *stage 1* of the loading process, the vertical compressive stress $P_{22} = P_{22}^*$ is applied to the particle aggregate in a stepwise fashion[2], using a total of $i_{vs}$ loading increments, with the subscript $vs$ designating "vertical stress". After initiating the displacement and rotation boundary conditions at the top $\partial V_t$ and bottom $\partial V_b$ boundaries, the vertical stress is incrementally updated and subsequently used to compute the displacement and rotation corrections at the left and right boundaries with expressions (31)-(32), and the displacement correction at the top boundary boundary with

$$\Delta u_{q,2} = g_a^t \Delta a_{q,2} \quad \text{with} \quad \Delta a_{q,2} = a_{q,2} - \bar{P}_{21} A_{q,1} - \bar{P}_{22}^* A_{q,2}. \qquad (47)$$

The expression above is derived from $(42)_2$, whereby during the incremental application of the vertical stress $\bar{P}_{22}^*$ the value of $\bar{P}_{21}$ is prescribed as zero, in order to avoid the initial development of a shear stress. After the particle aggregate has reached its equilibrium state under dynamic relaxation, the boundary forces and moments of the particles at the top, left and right boundaries are recorded and employed to compute the corresponding residuals. When all residuals are lower than the prescribed values of the corresponding tolerances, the iterative loop is stopped and the next vertical stress increment is applied. Otherwise, the iterative loop is entered again, until a converged solution is found. After the application of $i_{vs}$ increments the vertical stress has reached the desired value, and stage 1 of the loading process has completed.

During *stage 2* of the loading process, the horizontal shear deformation $\bar{F}_{12}$ is imposed on the particles at the top $\partial V_t$ and bottom $\partial V_b$ boundaries of the granular assembly, by displacing these in a stepwise manner using $i_{tot} - i_{vs}$ increments. The rotations of the particles at the top boundary are free, and the vertical displacement and rotation of the particles at the bottom boundary are fully constrained. In a similar way as explained above for stage 1, the boundary forces and moments in the relaxed equilibrium state are used to compute the displacement and rotation corrections at the periodic left and right boundaries $\partial V_l$ and $\partial V_r$, and at the top boundary boundary $\partial V_t$. However, the only difference is that in (47) the shear stress $\bar{P}_{21}$ here is not prescribed as zero, but is calculated from the homogenized response of the particle assembly using equation (23). After the dynamic relaxation procedure has completed, the residuals are computed in the same way as during stage 1, and compared against the corresponding tolerances. The iterative process is terminated when the convergence criterion is satisfied, after which the shear deformation is incremented and the response to the next loading step is computed. This procedure is continued until all loading increments $i_{tot}$ are applied.

---
[2]Instead of applying the vertical compressive stress by means of the first Piola-Kirchhoff stress $\bar{P}_{22}$, the Cauchy stress $\bar{\sigma}_{22}$ could have been used. The conversion of the Cauchy stress into the first Piola-Kirchhoff stress, which is the stress measure used in the numerical algorithm presented in Table 5, can straightforwardly be accomplished by using the inverse form of expression (30).

**ALGORITHM FOR MIXED BOUNDARY CONDITIONS**

---

**1. DEM simulation. Apply vertical stress. Increments** $0 \leq i \leq i_{vs}$

    **1.1** Apply boundary conditions

        **1.1.A** $q \in \partial V_t \Longrightarrow$ Free rotations $\boldsymbol{m}_q = \boldsymbol{0}$

        **1.1.B** $q \in \partial V_b \Longrightarrow$ Zero vertical displacements $x_{q,2} = 0$ and zero rotations $\mathbf{Q}_q = \mathbf{I}$

    **1.2** Update vertical stress $\bar{P}_{22} = \bar{P}_{22}^*$

    **1.3** Update particle configuration

        **1.3.A** $q \in \partial V_t \Longrightarrow$ Calculate particles displacement correction $\Delta u_{q,2}$ via (47), with $\bar{P}_{21} = 0$

        **1.3.B** $q \in \partial V_l \cup \partial V_r \Longrightarrow$

            Calculate particles displacement $\Delta \mathbf{u}_q$ and rotation $\Delta \theta_q$ corrections via (31)-(32)

    **1.4** Dynamic relaxation. Obtain boundary forces and moments.

    **1.5** Calculate residual(s)

        **1.5.A** $q \in \partial V_t \Longrightarrow r_a^t = \sqrt{\sum_{q \in \partial V_t} \Delta a_{q,2}^2 / \tilde{a}_q^2}$

        **1.5.B** $q \in \partial V_l \cup \partial V_r \Longrightarrow r_a^p = \sqrt{\sum_{q \in \partial V_l \cup \partial V_r} \Delta \mathbf{a}_q \cdot \Delta \mathbf{a}_q / \tilde{a}_q^2}$ and $r_m^p = \sqrt{\sum_{q \in \partial V_l \cup \partial V_r} (\Delta m_q / \tilde{m}_q)^2}$

    **1.6** Check for convergence: $r_a^t \leq \epsilon_a^t$ and $r_a^p \leq \epsilon_a^p$ and $r_m^p \leq \epsilon_m^p$

        **1.6.A** if converged $\Longrightarrow$ Save current configuration and go to 1 (next increment $i + 1$)

        **1.6.B** if not converged $\Longrightarrow$ Return to 1.3

---

**2. Apply horizontal shear deformation at fixed vertical stress. Increments** $i_{vs} < i \leq i_{tot}$

    **2.1** Apply updated macro-scale deformation and boundary conditions

        **2.1.A** $q \in \partial V_t \Longrightarrow$ Horizontal displacements $x_{q,1} = \bar{F}_{12} X_{q,2} + X_{q,1}$ and free rotations $\boldsymbol{m}_q = \boldsymbol{0}$

        **2.1.B** $q \in \partial V_b \Longrightarrow$ Zero vertical displacements $x_{q,2} = 0$ and zero rotations $\mathbf{Q}_q = \mathbf{I}$

            Horizontal displacements $x_{q,1} = \bar{F}_{12} X_{q,2} + X_{q,1}$

    **2.2** Dynamic relaxation. Obtain boundary forces and moments.

    **2.3** Update particle configuration

        **2.3.A** $q \in \partial V_t \Longrightarrow$ Calculate particles displacement correction $\Delta u_{q,2}$ via (47)

        **2.3.B** $q \in \partial V_l \cup \partial V_r \Longrightarrow$

            Calculate particles displacement $\Delta \mathbf{u}_q$ and rotation $\Delta \theta_q$ corrections via (31)-(32)

    **2.4** Dynamic relaxation. Obtain boundary forces and moments.

    **2.5** Calculate residual(s)

        **2.5.A** $q \in \partial V_t \Longrightarrow$ Refer to 1.5.A

        **2.5.B** $q \in \partial V_l \cup \partial V_r \Longrightarrow$ Refer to 1.5.B

    **2.6** Check for convergence: $r_a^t \leq \epsilon_a^t$ and $r_a^p \leq \epsilon_a^p$ and $r_m^p \leq \epsilon_m^p$

        **2.6.A** if converged $\Longrightarrow$ Save current configuration and go to 2 (next increment $i + 1$)

        **2.6.B** if not converged $\Longrightarrow$ Return to 2.3

---

Table 5: Algorithm for the application of the mixed boundary conditions. The loading process consists of stage 1, during which the vertical stress is incrementally applied, and stage 2, during which the horizontal shear deformation is incrementally imposed.

## 5.3 Computational results

The performance of the algorithm used for the implementation of the mixed boundary conditions is demonstrated by means of two DEM simulations of an irregular polydisperse packing of $449$ particles, with the particle radii taken randomly from a uniform size distribution with polydispersity $R_{max}/R_{min} = 1.5$, where the minimum radius equals $R_{min} = 0.8$ mm. The initial particle volume is $V = 1517$ mm$^2$, with the particle volume fraction of the packing being equal to $\mathfrak{v} = 0.849$, and the average particle coordination number as $3.55$. The two simulations consider different particle contact laws, namely the frictional contact law and the cohesive contact law described in Section 2.2.2. Assuming relatively hard particles, the normal and tangential contact stiffnesses for the frictional contact law are set as $k_n = 10^5$ N/mm and $k_s = 4 \cdot 10^4$ N/mm, respectively, and the friction coefficient equals $\mu = 0.6$. The normal contact stiffness $k_n^b$ and the tangential contact stiffness $k_s^b$ for the cohesive contact interaction are assumed to be equal to those of the frictional contact law, and the bending contact stiffness is taken as $k_\theta^b = 2 \cdot 10^4$ Nmm. The normal, shear and bending strengths have the values $f_n^{b,u} = 300$ N, $f_s^{b,u} = 60$ N and $m_\theta^{b,u} = 200$ Nmm, respectively. The density of the particles is $\rho = 10 \cdot 10^3$ kg/m$^2$. The macroscopic vertical (compressive) stress is $\bar{P}_{22}^* = -1.05 \cdot 10^6$ N/m, which is applied in $i_{vs} = 6$ increments. The total macroscopic shear deformation equals $\bar{F}_{12} = 0.2$, which is imposed on the particle aggregate in $i_{tot} - i_{vs} = 100$ increments. The translational and rotational damping factors used in the dynamic relaxation procedure are $\alpha = \beta = 0.7$, and the time increment equals $\Delta t = 10^{-5}$ s. The dimensionless values of the gain parameters are $g_a^t M_i / \Delta t^2 = g_a^p M_i / \Delta t^2 = 3 \cdot 10^4$ and $g_m^p M_i R_i^2 / \Delta t^2 = 6 \cdot 10^4$, and the corresponding tolerances are equal to $\epsilon_a^t = \epsilon_a^p = \epsilon_m^p = 2 \cdot 10^{-10}$. The above model parameters are summarized in Table 6.

| Parameter | Value | Unit |
| --- | --- | --- |
| Elastic normal stiffness $k_n = k_n^b$ | $1 \cdot 10^5$ | N/mm |
| Elastic tangential stiffness $k_s = k_s^b$ | $4 \cdot 10^4$ | N/mm |
| Elastic bending stiffness $k_\theta^b$ | $2 \cdot 10^4$ | Nmm |
| Friction coefficient $\mu$ | 0.6 | - |
| Cohesive normal strength $f_n^{b,u}$ | 300 | N |
| Cohesive tangential strength $f_s^{b,u}$ | 60 | N |
| Cohesive bending strength $m_\theta^{b,u}$ | 200 | Nmm |
| Density $\rho$ | $10 \cdot 10^3$ | kg/m$^2$ |
| Translational damping $\alpha$ | 0.7 | - |
| Rotational damping $\beta$ | 0.7 | - |
| Time increment $\Delta t$ | $10^{-5}$ | s |
| Tolerance force (P) $\epsilon_a^p$ | $2 \cdot 10^{-10}$ | - |
| Tolerance moment (P) $\epsilon_m^p$ | $2 \cdot 10^{-10}$ | - |
| Tolerance force (T) $\epsilon_a^t$ | $2 \cdot 10^{-10}$ | - |
| Gain force (P) $g_a^p M_i / \Delta t^2$ | $3 \cdot 10^4$ | - |
| Gain moment (P) $g_m^p M_i R_i^2 / \Delta t^2$ | $6 \cdot 10^4$ | - |
| Gain force (T) $g_a^t M_i / \Delta t^2$ | $3 \cdot 10^4$ | - |
| Tolerance dynamic relaxation $\text{tol}_E$ | $10^{-3}$ | - |

Table 6: Physical and algorithmic model parameters for the simulations with mixed boundary conditions.

Figure 11 shows the macroscopic response of the particle aggregates as a function of the applied shear deformation $\bar{F}_{12}$, with the dot-dashed and solid lines referring to packings with cohesive and frictional particle contact interactions, respectively. In Figure 11(a) the stress ratio $\bar{\sigma}_{12}/\bar{\sigma}_{22}$ is depicted, while Figure 11(b) illustrates the relative volumetric change $\det(\bar{F})$ (using the packing obtained after the application of the vertical stress as the reference state), and Figure 11(c) sketches the average particle rotation $\bar{\theta}$, in accordance with expression (40). Furthermore, in Figure 12 the particle configurations of the cohesive and frictional packings are plotted at four different deformation levels, namely (a) $\bar{F}_{12} = 0.002$, (b) $\bar{F}_{12} = 0.05$, (c) $\bar{F}_{12} = 0.1$ and (d) $\bar{F}_{12} = 0.015$. Here, the red lines between the particles represent cohesive contact forces, while the blue lines indicate frictional contact forces. It can be observed from Figure 11(a) that up to a deformation $\bar{F}_{12} = 0.02$ the cohesive and frictional packings show a similar response, whereby the stress increases approximately proportionally with deformation. Upon continuing deformation, in the frictional packing a large number of contacting particles meets the failure criterion (6) and starts to slide, such that the stress ratio $\bar{\sigma}_{12}/\bar{\sigma}_{22}$ reaches a maximum at $\bar{F}_{12} \approx 0.06$. The maximal stress ratio is accompanied by a volumetric increase of the particle structure, commonly referred to as "dilation", see Figure 11(b). After passing the peak value, the stress ratio for the frictional packing slightly drops in magnitude, which is caused by a substantial rolling of particles. The effect of particle rolling can be clearly observed from Figure 11(c), where at the onset of shear deformation the increase in average particle rotation is only mild, but steadily grows towards a more or less constant value at $\bar{F}_{12} = 0.07$, both for the frictional and cohesive packings. Note from Figure 11(c) that the initial value of the average particle rotation is due to the application of the vertical stress $\bar{P}_{22}^*$, and for the frictional packing appears to be somewhat larger than for the cohesive packing. For the cohesive packing the maximal value of the stress ratio $\bar{\sigma}_{12}/\bar{\sigma}_{22}$ is about 1.5 times larger than for the frictional packing, and is reached at $\bar{F}_{12} \approx 0.07$, see Figure 11(a). At this stage a significant number of particle bonds are broken, in correspondence with the failure criterion (8). With continuing deformation, the broken particle bonds of the cohesive packing become frictional, as indicated by the thick blue lines in Figure 12(c), whereby the stress ratio $\bar{\sigma}_{12}/\bar{\sigma}_{22}$ of the packing drops to a similar level as for the frictional packing, see Figure 11(a). Observe from Figures 12(c) and (d) that for the cohesive packing the frictional contacts indicated by the blue lines are established only along local force chains in the particle structure, whereby the rest of the contacts remain cohesive, as indicated by the red lines. This implies that the overall deformation of the packing towards the end of the loading process becomes governed by a localized failure zone, which indeed is associated to a strong softening behavior in the stress response, see Figure 11(a). Close to a shear deformation of $\bar{F}_{12} = 0.20$, both the stress ratio $\bar{\sigma}_{12}/\bar{\sigma}_{22}$ and the relative volumetric change $\det(\bar{F})$ of the frictional and cohesive packings become approximately constant, characterizing the occurrence of a so-called "critical state". The overall residual strength at the critical state is $\bar{\sigma}_{12}/\bar{\sigma}_{22} \approx 0.28$. This value is only about half of the value of 0.6 adopted for the local particle contact friction $\mu$, which can be explained from the fact that at the end of the deformation process the dilating particle structure, instead of sliding, is predominated by rolling of particles, see [7] for a more detailed discussion on this aspect.

As a final note, it is mentioned that the contact moments in the cohesive packing determine about 10% of the total potential energy. The reason that this contribution does not become manifest in the stress ratio $\bar{\sigma}_{12}/\bar{\sigma}_{22}$ depicted in Figure 11(a), is because the Cauchy stress is determined by contact forces only, see expressions (23) and (30). As already mentioned in Section 2.4, the overall effect by the contact moments on the particle aggregate can be accounted for through the introduction of a couple stress, but this stress measure is left out of consideration in the present study.

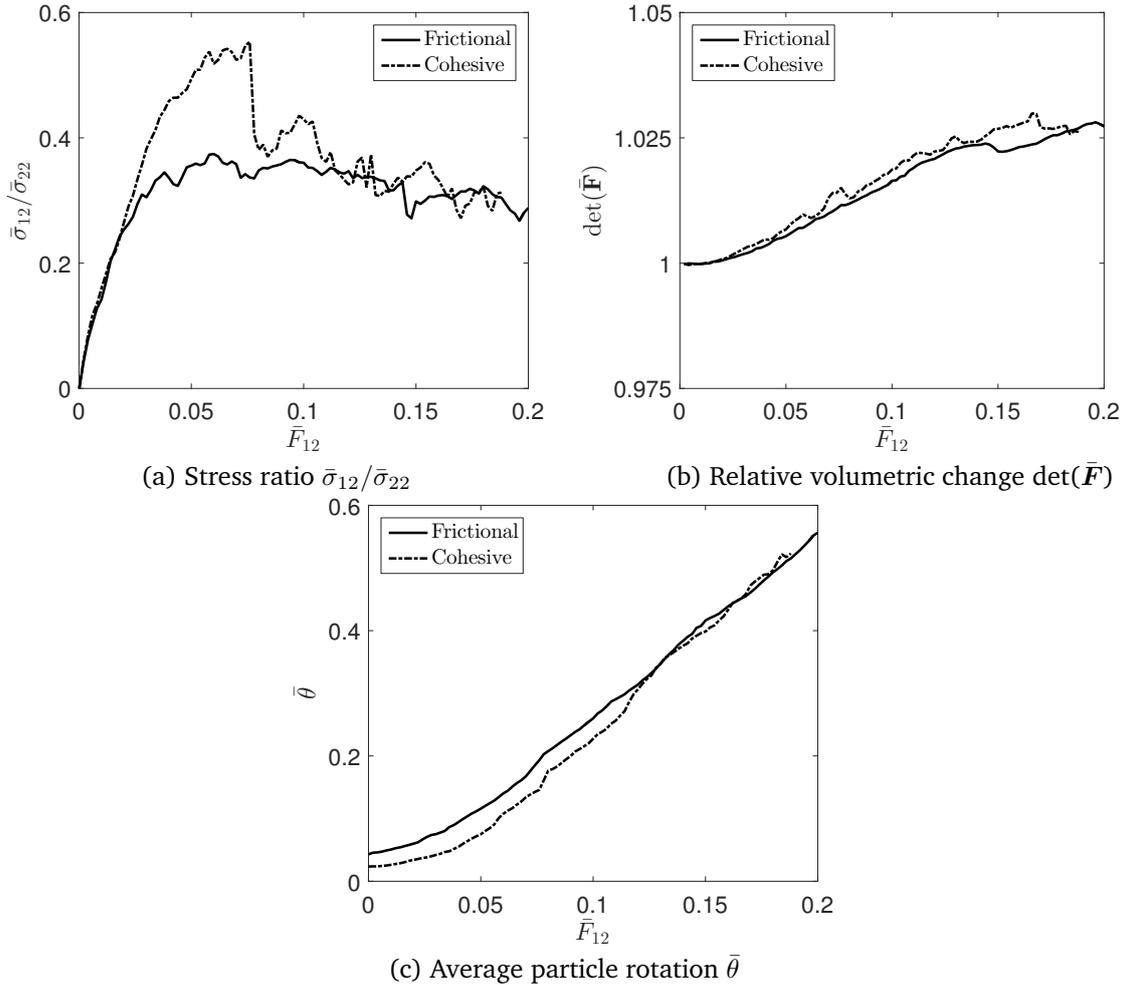

Figure 11: Macroscopic response of an infinite granular layer subjected to a vertical compressive stress $\bar{P}_{22} = \bar{P}_{22}^*$ and a horizontal shear deformation $\bar{F}_{12}$. (a) Stress ratio $\bar{\sigma}_{12}/\bar{\sigma}_{22}$, (b) relative volumetric change $\det(\bar{F})$, and (c) average particle rotation $\bar{\theta}$, all plotted versus the applied shear deformation $\bar{F}_{12}$ for cohesive (dot-dashed line) and frictional (solid line) packings.

# 6 Conclusions

Novel numerical algorithms have been presented for the implementation of three types of classical boundary conditions for a particle aggregate. The micro-scale boundary conditions are formulated within the discrete element method using large deformation theory, and, along the lines of [16], are imposed on a frame of boundary particles of the particle packing, in accordance with i) a homogeneous deformation and zero particle rotation (D), ii) a periodic particle displacement and rotation (P), and iii) a uniform particle force and free particle rotation (T). The algorithms can be straightforwardly combined with commercial discrete element codes, thereby enabling the determination of the solution of boundary-value problems at the micro-scale only, or at multiple scales via a micro-to-macro coupling with a finite element model. The performance of the algorithms has been tested by means of discrete element method simulations on regular monodisperse packings and irregular polydisperse packings composed of frictional particles, which were subjected to various loading paths. The simulations provide responses with the typical stiff and soft bounds for the (D) and (T) boundary conditions, respectively, and illustrate for the (P) boundary condition a relatively fast convergence of the apparent macroscopic properties under an increasing packing size. Finally, a homogenization framework has been presented for the formulation of mixed (D)-(P)-(T) boundary conditions that satisfy the Hill-Mandel micro-heterogeneity condition on energy consistency at the micro- and macro-scales of the granular system. The numerical algorithm for mixed boundary conditions has been developed and tested for the case of an infinite layer subjected to a vertical compressive stress and a horizontal shear deformation, whereby the response computed for a layer of cohesive particles is compared against that for a layer of frictional particles. The results illustrate that the failure response for both contact laws is characterized by the development of a dilated particle structure, which at large deformation gradually turns into a critical state with an approximately constant residual strength and specific volume. The application of the present algorithms for multi-scale FEM-DEM analyses on granular systems with a large number of particles, and their extension towards a dynamics homogenization framework, are topics for future studies.

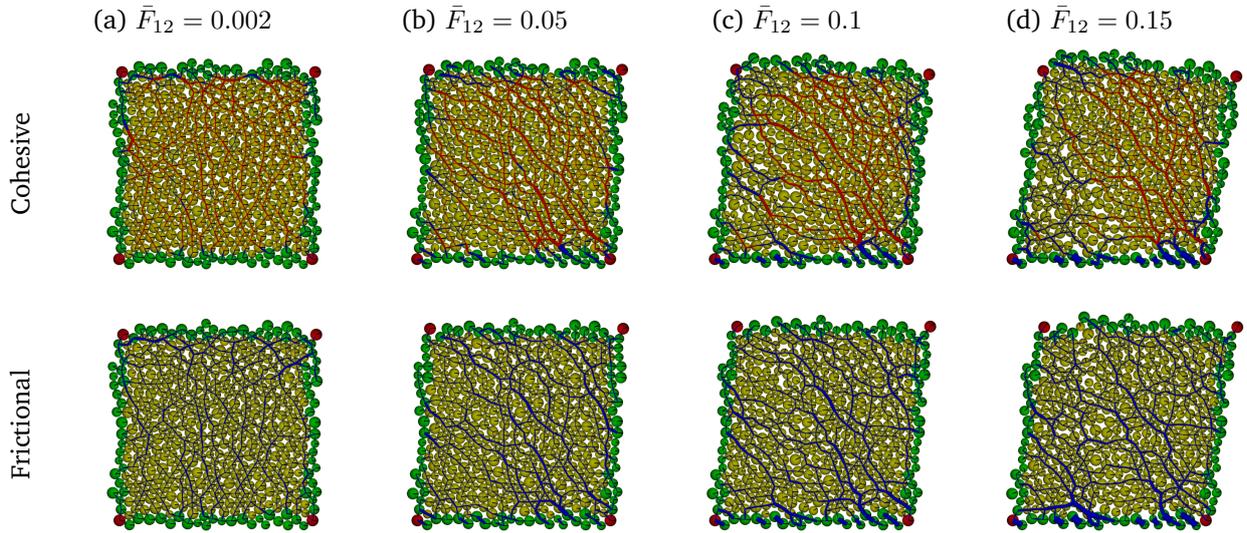

Figure 12: Deformed configurations of a packing with 449 particles corresponding to (a) $\bar{F}_{12} = 0.002$, (b) $\bar{F}_{12} = 0.05$, (c) $\bar{F}_{12} = 0.1$ and (d) $\bar{F}_{12} = 0.15$, for cohesive (top) and frictional (bottom) particle contact interactions. The networks of cohesive and frictional forces acting between particles are indicated by the red and blue lines, respectively.

# 7 Acknowledgments

The authors thank Dr. Ning Guo from Carleton University, Ottawa, Canada, and Dr. Lutz Gross from the University of Queensland, Brisbane, Australia, for the helpful discussions on the use of Escript, and Prof. Stefan Luding from the University of Twente, Netherlands for the useful interactions on DEM modeling. The feedback provided by Dr. Steffen Abe and Dr. Dion Weatherley from the University of Queensland, and by Joran Jessurun from the Eindhoven University of Technology, Netherlands, on numerical implementation issues within the ESyS-Particle code, is appreciated. The financial support of J.L. by the China Scholarship Council is gratefully acknowledged.